\newcommand{\eqnn}[1]{\begin{eqnarray*}#1\end{eqnarray*}}
\newcommand{\eqnl}[2]{\par\parbox{12cm}
{\begin{eqnarray*}#1\end{eqnarray*}}\hfill
\parbox{1cm}{\begin{eqnarray}\label{#2}\end{eqnarray}}\break}

\newcommand{\eqn}[1]{\begin{eqnarray*}#1\end{eqnarray*}}
\newcommand{\eqngr}[2]{\begin{eqnarray*}#1\\#2\end{eqnarray*}}
\newcommand{\eqngrlb}[3]{\par\parbox{11cm}
{\begin{eqnarray}\fbox{$\displaystyle#1\\#2$}\end{eqnarray}}\hfill
\parbox{1cm}{\begin{eqnarray}\label{#3}\end{\eqnarray}}\break}
\newcommand{\eqngrl}[3]{\par\parbox{12cm}
{\begin{eqnarray*}#1\\#2\end{eqnarray*}}\hfill
\parbox{1cm}{\begin{eqnarray}\label{#3}\end{eqnarray}}\break}

\newcommand{\eqngrr}[3]{\begin{eqnarray*}#1\\#2\\#3\end{eqnarray*}}
\newcommand{\eqngrrl}[4]{\par\parbox{12cm}
{\begin{eqnarray*}#1\\#2\\#3\end{eqnarray*}}\hfill
\parbox{1cm}{\begin{eqnarray}\label{#4}\end{eqnarray}}\break}

\newcommand{\refs}[1]{(\ref{#1})}

\def\dmu{\pa_\mu}
\def\ilap{\frac{1}{\triangle}}

\def\lap{\triangle}
\def\es{\!=\!}
\def\pa{\partial}
\def\Box{\Delta}
\def\al{\alpha}

\def\lam{\lambda}
\def\si{\sigma}

\def\ov{\over}
\def\di{D\!\!\!\!\slash\,}
\def\fdi{\partial\!\!\!\slash\,\,}
\def\ha{{1\over 2}}
\def\tr{\,{\rm tr}\,}
\def\pr{\prime}
\def\gam{\gamma}
\def\gamfive{\gamma_5}
\def\om{\omega}

\def\cd{{\cal D}}
\def\cs{{\cal S}}
\def\R{{\cal R}}
\def\ra{\rangle}
\def\la{\langle}
\def\gf{\gamma_5}
\def\pan{\par\noindent}
\def\de{\delta}
\def\pd{\psi^{\dagger}}
\def\mtxt#1{\quad\hbox{{#1}}\quad}
\def\be{\bar\eta}

\font \fivesans               = cmss10 at 5pt

\font \sevensans              = cmss10 at 7pt

\font \tensans                = cmss10

\newfam\sansfam
\textfont\sansfam=\tensans\scriptfont\sansfam=\sevensans
\scriptscriptfont\sansfam=\fivesans
\def\sans{\fam\sansfam\tensans}

\def\bbb1{{\rm 1\!1}}

\def\bbbc{{\mathchoice {\setbox0=\hbox{$\displaystyle\rm C$}\hbox{\hbox
to0pt{\kern0.4\wd0\vrule height0.9\ht0\hss}\box0}}
{\setbox0=\hbox{$\textstyle\rm C$}\hbox{\hbox
to0pt{\kern0.4\wd0\vrule height0.9\ht0\hss}\box0}}
{\setbox0=\hbox{$\scriptstyle\rm C$}\hbox{\hbox
to0pt{\kern0.4\wd0\vrule height0.9\ht0\hss}\box0}}
{\setbox0=\hbox{$\scriptscriptstyle\rm C$}\hbox{\hbox
to0pt{\kern0.4\wd0\vrule height0.9\ht0\hss}\box0}}}}

\def\bbbq{{\mathchoice {\setbox0=\hbox{$\displaystyle\rm Q$}\hbox{\raise
0.15\ht0\hbox to0pt{\kern0.4\wd0\vrule height0.8\ht0\hss}\box0}}
{\setbox0=\hbox{$\textstyle\rm Q$}\hbox{\raise
0.15\ht0\hbox to0pt{\kern0.4\wd0\vrule height0.8\ht0\hss}\box0}}
{\setbox0=\hbox{$\scriptstyle\rm Q$}\hbox{\raise
0.15\ht0\hbox to0pt{\kern0.4\wd0\vrule height0.7\ht0\hss}\box0}}
{\setbox0=\hbox{$\scriptscriptstyle\rm Q$}\hbox{\raise
0.15\ht0\hbox to0pt{\kern0.4\wd0\vrule height0.7\ht0\hss}\box0}}}}

\def\bbbz{{\mathchoice {\hbox{$\sans\textstyle Z\kern-0.4em Z$}}
{\hbox{$\sans\textstyle Z\kern-0.4em Z$}}
{\hbox{$\sans\scriptstyle Z\kern-0.3em Z$}}
{\hbox{$\sans\scriptscriptstyle Z\kern-0.2em Z$}}}}
\def\qed{\ifmmode\sq\else{\unskip\nobreak\hfil
\penalty50\hskip1em\null\nobreak\hfil\sq
\parfillskip=0pt\finalhyphendemerits=0\endgraf}\fi}

\newif\iffigs\figsfalse
\iffigs
 \input epsf
 \def\PSGraphic#1#2{
 \centerline{\epsfbox{#1}}  \vskip 10pt} 
\fi

\documentstyle[12pt]{article}
\begin{document}


\hsize=6in
\titlepage
\begin{flushright} DIAS-STP-95-19 \\  FSUJ TPI 4/95\\ July 1995
\end{flushright}
\begin{center}\Large{{ Generalized Thirring Models }} \end{center}
\vspace{1cm}
\begin{center} I. Sachs$\;^\dagger$ and A. Wipf$\;^\ddagger$\\
\end{center}
\vspace{1cm}
\begin{center}
$^\dagger\;$Dublin Institute for Advanced Studies\\
10 Burlington Road, D 4, Ireland\\
$^\ddagger\;$Theoretisch-Physikalisches Institut,
Friedrich-Schiller-Universit\"at,
\\Max-Wien-Platz 1, 07743 Jena, Germany\\
\end{center}
\vspace{3cm}


\begin{abstract}The Thirring model and various generalizations of it are
analyzed in detail. The four-Fermi interaction  modifies the equation of state.
Chemical potentials and twisted boundary conditions both result
in complex fermionic determinants which are analyzed. The non-minimal
coupling to gravity does deform the conformal algebra which in particular
contains the minimal models. We compute the central charges,
conformal weights and finite size effects.\hfill\break
For the gauged model we derive the partition functions and the explicit
expression for the chiral condensate at finite temperature and curvature.
The Bosonization in compact curved space-times is also investigated.
\end{abstract}
\def\sy{\scriptscriptstyle}<
\newpage

\section{Introduction}
The response of physical systems to a change of external conditions
is of eminent importance in physics. In particular the dependence
of expectation values on temperature, the particle density,
the space region, the imposed boundary conditions or external fields
has been widely studied \cite{chh4tdr84}. Nevertheless, many properties of
such systems are poorly understood. The massless Thirring model
\cite{chh4tt58}, which is among the simplest interacting field theories,
has already led to considerable confusion about its thermodynamic
properties in the literature \cite{chh4ty87,chh4tfp88,chh4tdv89}.
The reason is two-fold: Firstly, the computation of the fermionic determinant
in the presence of a chemical potential and/or non-trivial boundary conditions
is delicate, because the eigenvalues of the Dirac operator are generically
complex. In section $3.1$ we propose a regularization scheme via
analytic continuation. We argue that the so-obtained determinant, which
differs from previous results \cite{chh4tfp88}, leads to the correct equation
of state.\hfill\break
The second complication originates in the infrared-sector. An elegant
infrared regularization, which is particularly well suited for the study of
thermodynamic properties, is to quantize the model on a torus. Harmonic
contributions to the current arise then naturally and taking them into
account turns out to be crucial for a correct quantization. In particular the
so-obtained results differ from those gotten earlier \cite{chh4ty87}
using bosonization. This is explained in section $3.2$.\par

On another front there has been much effort to quantize self-interacting
field theories in a background gravitational field \cite{chh4tbd82}.
For example, one is interested whether a black hole still emits thermal
radiation when self-interaction is included. Due to general arguments by
Gibbons and Perry \cite{chh4tgp78} this question is intimately connected
with the universality
of the second law of thermodynamics. The Thirring model (including the gauged
version of it) is still solvable in curved space-time and we can study its
properties in a background gravitational field. This provides
us in particular with an elegant approach to the study of its conformal
structure:
Correlation functions with current- and stress-tensor insertions, which
are gotten by functional differentiation with respect to the gauge- and
gravitational fields, contain the necessary information to
characterize the underlying symmetry algebras. To
familiarize the reader with our approach we first rederive the conformal
structure of the original Thirring model in section $3.3$. We then show
how a non-minimal coupling to gravity leads in a natural way to a
modification of the conformal structure. In particular, very much as for a
free scalar field the central charge in the fermionic formulation of the
Thirring model is not unique. Furthermore, we find that the equivalence
between finite size
scaling and central charge of the Virasoro algebra holds only
for a particular treatment of the zero-mode sector in which a charge
at infinity is generated automatically. This charge combines in a
non-trivial way with the Weyl-anomaly of the determinant of the
fluctuation operators to reconcile the equivalence of the finite size
scaling and the central charge. For certain values of the non-minimal
coupling we obtain minimal models from interacting fermions.
This is the subject of sections 3.4 and 3.5.\par
The gauged Thirring model, which contains the Schwinger model ($QED_2$) as a
particular limit, is no longer conformally invariant but has a mass gap:
The 'photon' acquires a mass $m_\gamma^2=e^2/(\pi+\ha  g_{{\sy 2}}^2)$ via
the Schwinger mechanism. It possesses a non-trivial vacuum structure
which promotes it to an
attractive toy model to mimic the complex vacuum structure in $4$-d gauge
theories.  From our experience with the Schwinger model \cite{chh4tsw92},
which is supposed to share certain aspects with one-flavour $QCD$
\cite{chh4tls92}, we expect that gauge fields with winding numbers are
responsible for the non-vanishing chiral condensate and in particular
its temperature dependence. Configurations with windings, so called
instantons, exist only for finite volumes and minimize the Euclidean action.
They lead to chirality violating vacuum expectation values. For example,
a non-zero chiral condensate develops which only for high temperature
and large curvature vanishes exponentially.\par
Since for particular choices of the coupling constants the
model reduces to well known and well studied exactly soluble
models there are many earlier works which are related
to ours. Some of them concentrated more on the gauge sector
and investigated the renormalization of the electric charge
by the four-Fermi interaction
\cite{chh4tj64}  or the non-trivial vacuum structure in the
Schwinger model \cite{chh4tsw92,chh4tj88}. Others concentrated
on the un-gauged conformal sector. Freedman and Pilch calculated the
partition function of the un-gauged Thirring model on arbitrary
Riemann surfaces \cite{chh4tfp88}. We do not agree with their
result and in particular show that there is no
holomorphic factorization for general
fermionic boundary conditions. Also we deviate from Destri and deVega
\cite{chh4tdv89}
which investigated the un-gauged model on the cylinder
with twisted boundary conditions.
We comment on these discrepancies in section $3.2$.\hfill\break
Section $2$ contains introductory
material and in particular the classical structure of the model.\par
Other papers are dealing with
different aspects of certain limiting cases of the model considered here.
In particular in \cite{chh4ty87}, the thermodynamics of the Thirring model
has been studied and the Hawking radiation
has been derived in \cite{chh4tbd78}. The equivalence of the massive Thirring
model and the Sine-Gordon model in curved space has been shown in \cite{Eboli}.
Partition functions for scalar fields
with twisted boundary conditions have been computed in \cite{chh4tg94} and
more recently in \cite{denjoe}.\par

\newpage
\section{Classical Theory}

The gauged Thirring model in curved space-time has the Lagrangian
\eqnl{
{\cal L}_{\rm Thir}[A_\mu,\bar\psi,\psi]=\bar\psi i \gamma^\mu\nabla_\mu
\psi-{g^2\ov 4}
j^{\mu}j_{\mu}- {1\over 4} F_{\mu \nu} F^{\mu \nu}\mtxt{,}
j^\mu=\bar\psi\gamma^\mu\psi\;,}{t13}
where the gamma-matrices in curved space-time are related to the ones
in Minkowski space-time
as $ \gamma^\mu\es e^\mu_{\,a}\hat \gamma^a$, $\nabla_\mu
\es\pa_\mu\!+i\omega_\mu\!-ieA_\mu$ is the coordinate- and
gauge covariant derivative and
$F_{\mu\nu}$ is the electromagnetic field strength.
The gravitational field
$g_{\mu\nu}$ (or rather the $2$-bein $e^a_{\,\mu}$, since the
theory contains fermions) is treated as classical background field,
whereas the 'photons' $A_\mu$ and fermions $\psi$ will be quantized.
\hfill\break
The classical theory is invariant
under $U(1)$ gauge- and axial transformations and correspondingly
possesses conserved vector and axial-vector currents
\eqnl{
j^\mu\mtxt{and}j^{{\sy 5}\mu}\es
\bar\psi\gamma^\mu\gamma_{{\sy 5}}\psi\es\eta^\mu_{\;\;\nu}j^\nu.}{t3}
Here $\eta_{\mu\nu}\es\sqrt{-g}\,\epsilon_{\mu\nu}$ denotes the
totally antisymmetric tensor.
In fact, the conservation laws together with the relation \refs{t3} between the
vector- and axial currents imply that the currents are free
fields
\eqnl{
\nabla^2 j^\mu=\;\nabla^2 j^{{\sy 5}\mu}=\;0\;,}{t8}
which is the reason that accounts for the solubility
of the model \cite{chh4tk68}, even in the presence of gauge- and
gravitational fields. Of course, for any gauge
invariant regularization the axial current possesses an anomalous
divergence in the quantized model.
Thus the normal $U_A(1)$ Ward identities in the un-gauged Thirring
model \cite{chh4tj64} become anomalous when the fermions couple to a gauge
field.\par
The solution of the equations of motion is most easily presented by introducing
auxiliary scalar- and pseudo-scalar fields, in terms of which the action takes
the form
\eqngrl{
{\cal S}&=&\int\sqrt{-g}\Big[-{1\over4}
F_{\mu\nu}F^{\mu\nu}+i\bar\psi \gamma^\mu(\nabla_\mu- ig_{{\sy 1}}\pa_\mu
\lam+
i g_{{\sy 2}}\eta_\mu^{\;\;\nu}\pa_\nu\phi )\psi\Big]}
{&&\qquad\qquad+g^{\mu\nu}(\pa_\mu\phi\pa_\nu\phi+
\pa_\mu\lam\pa_\nu\lam )\Big].}{t1}
Note that for later use we have allowed for different couplings of the
fermionic currents to the scalar- and pseudo-scalar auxiliary fields
$\lam$ and $\phi$, respectively. The
original Thirring model is recovered for $g_{{\sy 1}}\es g_{{\sy 2}}\es g$,
since then
\eqnl{
{\cal L}=-{1\over4}F^2+i\bar\psi \gamma^\mu\nabla_\mu\psi
+gj^\mu B_\mu+g^{\mu\nu}B_\mu B_\nu,\quad
B_\mu=\pa_\mu\lam-\eta_\mu^{\;\;\nu}\pa_\nu\phi}{t1a}
is classically and quantum-mechanically equivalent to
\refs{t13}, after elimination of the multiplier field $B_\mu$.\par
\def\F{g_{{\sy 1}}\lam}
\def\G{ g_{{\sy 2}}\phi}
By decomposing the gauge field similarly as the $B_\mu$-field as
\eqnl{
A_\mu=\dmu\al-\eta_{\mu \nu} \pa ^\nu   \varphi,
\quad\mtxt{so that} F_{01}=\sqrt{-g}\nabla^2   \varphi,}{t4}
and choosing isothermal coordinates for which
$g_{\mu\nu}\es e^{2\si}\eta_{\mu\nu}$, the generalized Dirac operator reads
\eqngrl{
\di&=&\;e^{i F-i\gamfive G-{3\over 2}\si}\;\pa\!\!\!\slash \;
e^{-i F-i\gamfive G+\ha\si},\mtxt{where}}
{F&=& g_{{\sy 1}}\lam+e\;\al\mtxt{and}G=  g_{{\sy 2}}\phi+e\;\varphi\;.}{t5}
Hence, if $\psi_{\sy 0} (x)$ solves the free Dirac equation
in flat Minkowski space time, then
\eqnl{\psi (x) \equiv e^{ iF + i \gamfive G - \ha  \si}\psi_{\sy 0}}{t6}
solves the Dirac equation of the interacting theory on curved space-time.
The vector currents are related as
\eqnn{
j^\mu=\bar\psi\gam^\mu\psi=\bar\psi_0\hat\gam^\mu\psi_0e^{-2\sigma}
\equiv {1\ov \sqrt{-g}}j^\mu_0.}
The same relation holds for the axial vector current.\par
Diffeomorphism invariance then leads covariantly conserved energy-moment
um tensor
\eqnl{
T^{\mu\nu}\equiv-{2\ov\sqrt{g}}{\de\cs\ov\de g_{\mu\nu}}.}{t11}
Applying the variational identities in Appendix A one
obtains after a lengthy but straightforward computation
\begin{eqnarray}
T^{\mu\nu}&=&\;{1\over 4}g^{\mu\nu}F^{\si\rho}F_{\si\rho}
-F^{\si \nu}F_{\si}^{\;\;\mu}+{i\over 2}\bigl [ \bar\psi\gamma^{(\mu}
D^{\nu)}\psi -(D^{(\mu}\bar\psi) \gamma^{\nu)}\psi\bigr ]\nonumber\\
&&\quad+2 \nabla^\mu\phi\nabla^\nu\phi - g^{\mu\nu} \nabla^\al\phi\nabla_\al
\phi
\quad + \quad (\phi \leftrightarrow \lam )\label{t12}\\
&&\quad+\ha j^{\mu}\;(g_{{\sy 1}}\nabla^{\nu}\lam- g_{{\sy 2}}\eta^{\nu \al}
\nabla_\al
\phi)\quad+\quad (\mu \leftrightarrow\nu)\nonumber\\
&&\quad +  g_{{\sy 2}}  g^{\mu\nu} j^\al\eta_{\al\beta}\nabla^\beta\phi-2
g_{{\sy 2}}
j^\al\eta_\al\,^{(\mu}\nabla^{\nu)}\phi \; ,\nonumber
\end{eqnarray}
where we have introduced the symmetrization $A^{(\mu }B^{\nu)}\es
\ha (A^\mu B^\nu + A^\nu B^\mu)$. The first two lines are just
the energy momentum of the electromagnetic field, charged
fermions and free neutral (pseudo-) scalars. The remaining terms
reflect the interaction between the fermionic and auxiliary
fields. On shell $T^{\mu\nu}$ is conserved as required by general
covariance. Using the field equations for $\psi$ and $\lam$
its trace reads
\eqnl{
T^{\mu}_{\;\mu}=\;-\ha F^{\si\rho}F_{\si\rho}\;.}{t12a}
In particular for $A_\mu \es 0$ it
vanishes, and the theory becomes Weyl-invariant.\par

\def\dr{\overleftarrow{\delta}}
\def\dl{\overrightarrow{\delta}}
\paragraph{Symplectic structure:} In the presence of both
fermions and bosons it is convenient to exploit the graded
Poisson structure \cite{chh4tc76}
\eqnn{
\{A(x),B(y)\}\equiv \sum_O\int dz^1
\Bigl(
{A(x)\dr\ov\de O(z)}{\dl B(y)\ov\de\pi_O(z)}\mp
{A(x)\dr\ov\de \pi_O(z)}{\dl B(y)\ov\de O(z)}\Bigr)
\Big|_{x^0=y^0}\;.}
The sum is over all fundamental fields $O(x)$ in the theory .
The sign is minus if one or both of the fields $A$ and $B$ are
bosonic (even) and it is plus if both are fermionic (odd) fields.
The momentum densities $\pi_O(x)$ conjugate to the $O$-fields are
given by functional left-derivatives.
A simple calculation yields the following momenta
\eqnn{
\pi_\psi=-i \psi^\dagger,\quad
\pi_\phi=  g_{{\sy 2}} j^{\sy 5}_0+2 \pa_0\phi\mtxt{and}
\pi_\lam= g_{{\sy 1}} j_0+2\pa_0\lam.}\par
In the following sections we are lead to consider the {\it Euclidean
version} of the model. Then one must replace the Lorentzian
$ \gamma^\mu,g_{\mu\nu}$ and $\om _\mu$ by their Euclidean
counterparts. For example, with our conventions
the relation \refs{t3} becomes
\eqnn{
j^{{\sy 5}\mu}=-i\eta^\mu_{\;\;\nu}\,j^\nu}
and the generalized Dirac operator in Euclidean space-time becomes
\eqnl{
\di=\; e^{-2\sigma}e^{f^\dagger}\;
\fdi\;e^{f},\mtxt{where}f=-iF+\gamma_5G+\ha\sigma}{scal1}
(see \refs{t5} for the definition of $F$ and $G$), instead of \refs{t5}.
Also, to recover the Euclidean Thirring
model as particular limit of \refs{t1} we must set $ g_{{\sy 2}}^2\es
g_{{\sy 1}}^2\es g^2$.

\section{Thermodynamic- and conformal properties}

In this section we analyze the quantum theory corresponding to the classical
action \refs{t1} without gauge fields, in flat and curved
space-time. The gauged model is then considered in the next section.
Here we calculate the
{\it partition function, ground state energy, equation of state}
and determine the {\it conformal structure} of the un-gauged model.\par
To allow for a non-vanishing U(1)-charge we couple this
conserved charge to a chemical potential $\mu$. For the finite
temperature model the imaginary
time must vary from zero to the inverse temperature $\beta$
and the bosonic and fermionic fields must obey periodic and anti-periodic
boundary conditions, respectively. We enclose the system in a
spatial box with length $L$ to avoid infrared divergences. \par
We shall determine the dependence of the partition function and correlators on
the metric. This provides us with an alternative approach
to the conformal structure and its relation to finite size-effects.
Also, it enables us to study the effect of non-minimal
coupling to gravity in section $3.3$. Hence we allow for an arbitrary metric
or $2$-bein $e_{\mu a}$ with Euclidean signature.
We can choose (quasi) isothermal coordinates and a Lorentz frame such
that
\eqngrl
{e_{\mu a}&=& e^\si \,\hat e_{\mu a}\equiv
e^\si \pmatrix{\tau_0&\tau_1\cr 0 & 1}}
{g_{\mu\nu}&=&e^{2\si }\,\hat g_{\mu\nu}\equiv
e^{2\si }\pmatrix{\vert\tau\vert^2 & \tau_1\cr \tau_1&1},\qquad
\sqrt{g}=e^{2\sigma}\tau_0,}{t517}
where $\tau\es\tau_1+i\tau_0$ is the Teichmueller parameter
and $\si $ the gravitational Liouville field. Space-time
is then a square of length $L$ and has volume $V\es\int_0^L d^2x
\sqrt{g}$. We allow for the general twisted boundary conditions
for the fermions
\eqngrl{
\psi(x^0+L,x^1)&=&-e^{2\pi i(\al_0+\beta_0\gf)}\psi(x^0,x^1)}
{\psi(x^0,x^1+L)&=&-e^{2\pi i(\al_1+\beta_1\gf)}\psi(x^0,x^1).}{t518}
The parameters $\al_i$ and $\beta_i$ represent {\it vectorial
and chiral twists}, respectively. We could allow for twisted
boundary conditions for the (pseudo) scalars as well \cite{chh4tg94,denjoe},
e.g.
$\phi(x^0\!+\!nL,x^1\!+\!mL)=\phi(x^1,x^0)+2\pi(m\!+\!n)$.
However, to recover the Thirring model for equal couplings we
must assume that these fields are periodic. For
$\si \es 0$, $\tau\es i\beta/L$ in which case $V=\beta L$, and for
$\al_0\es \beta_0\es0$ the
partition function has the usual thermodynamical
interpretation. Its logarithm is proportional to the free energy
at temperature $T\es 1/\beta$.

\subsection{Fermionic Generating Functional}
Twisted boundary conditions as in \refs{t518} require some care
in the fermionic path integral. The subtleties are not related to the
unavoidable ultra-violet
divergences but to the transition from Minkowski- to Euclidean
space-time. To see that more clearly let ${\cal{S^\pm}}$
denote the space of fermionic fields in {\it Minkowski space-time}
with chirality $\pm1$. Since both the commutation relations
and the action do not connect ${\cal{S^+}}$ and ${\cal{S^-}}$ we
can consistently impose different boundary conditions on
${\cal{S^+}}$ and ${\cal{S^-}}$. On the other hand, in the
{\it Euclidean path-integral} for the generating
functional
\eqnl{
Z_F[\eta,\bar\eta]=\int \cd \pd\cd \psi \;e^{\int\sqrt{g}\,\pd
i\di \psi+\int \sqrt{g} \,\big(\bar\eta\psi+\pd\eta\big)},}{t19}
the Dirac operator
\eqnn{\di =\pmatrix{0&D_-\cr D_+&0\cr}}
exchanges the two chiral components of $\psi$, i.e.
$\di :{\cal{S^\pm}}\rightarrow {\cal{S^\mp}}$. Thus, in
contrast to the situation in Minkowski space the two chiral
sectors are related in the action. Of course, the eigenvalue
problem for $i\di$ is then not well defined. This
is the origin of the ambiguity in the definition of
the determinant. It is related to the ambiguities one
encounters when one quantizes chiral fermions \cite{chh4taw83}. Here we
reformulate this problem in such a way that the determinant with chiral
twists ($\beta\!\neq\! 0$) can be obtained by analytic continuation.
The resulting determinants do not
factorize into (anti-) holomorphic pieces.
In appendix B we give further arguments in favour of our result
by calculating the determinants in a different way.
\par

Let us now study the generating functional for fermions in an external
gravitational and auxiliary field.
For that we observe that on the torus we must add a harmonic piece
to the auxiliary fields to which the fermionic current couples
in \refs{t1}. More precisely, in the Hodge-decomposition of
$B_\mu$ in \refs{t1a} contains a harmonic piece,
\eqnl{
B_\mu=\pa_\mu\lam-\eta_\mu^{\;\;\nu}\pa_\nu\phi+{2\pi\ov L}h_\mu
\mtxt{with}\nabla^\mu h_\mu=h_{[\mu;\nu]}=0.}{help}
More generally, allowing for arbitrary couplings of the various terms
in \refs{help} to the fermionic current, we are led to add a term
\eqnn{
{2\pi \ov L}g_{{\sy 0}}\int \sqrt{g}\,h_\mu j^\mu+({2\pi \ov L})^2\int
\sqrt{g}\, h_\mu h^\mu}
to the action \refs{t1}. Note, that in isothermal coordinates, for which
the metric has the form \refs{t517}, the harmonics $h_\mu$ are constant.
The constant $h_\mu$ couple to the harmonic part
of the current and are needed to recover the Thirring model in the
limit $g_{{\sy 0}}^2\es g_{{\sy 1}}^2\es  g_{{\sy 2}}^2$. Also, we shall see
that the harmonic degrees of freedom are essential to obtain the
correct thermodynamic potential.\hfill\break
Finally we introduce a {\it chemical potential} for the conserved
$U(1)$ charge. In the Euclidean functional approach
this is equivalent to coupling the fermions to a constant
imaginary gauge potential $A_0$ \cite{chh4ta87}.\hfill\break
As a consequence of the above observations the scaling formula \refs{scal1}
(recall, that $F\es g_{\sy 1}\lam$ and $G\es g_{\sy 2}\phi$ when
the electromagnetic interaction is switched off) is modified to
\eqngrl{
\di&=&e^{-2\sigma}e^{f^\dagger}\,\hat\di\,
e^{f},\quad\mtxt{where}\qquad f=-ig_{\sy 1}\lam+\gam_{\sy
5}g_{\sy2}\phi+\ha\sigma}
{\hat\di&=&\gamma^\mu\Big(\pa_\mu+i\hat\omega_\mu
-{2\pi i\ov L}[g_{{\sy 0}} h_\mu+\mu_\mu]\Big)
\mtxt{and} \mu_\mu=-i{\tau_0L\ov2\pi}\mu\;\delta_{\mu 0}.}{t28}
This scaling property will enable us to relate the fermionic
determinants and Green's functions of $\di$ and $\hat\di$. The spin connection
$\hat\omega$ in \refs{t28} vanishes
for our choice of the reference
zweibein. The dependence of $\hat\di$ on the chemical potential
$\mu$ and the constant harmonic field $h_\mu$ cannot be gotten
by the anomaly equation \cite{chh4tbv89}. It must
be computed by direct methods. For this we expand the fermionic
field in a orthonormal basis of the Hilbert space
\eqngrl{
\psi(x) &=& \sum_n a_n\psi_{n+}(x)+ \sum_n b_n\psi_{n-}(x)}
{\psi^\dagger(x) &=& \sum_n \bar a_n\chi_{n+}^\dagger (x)+\sum_n
\bar b_n\chi_{n-}^\dagger(x),}{ba1}
where $a_n,b_n,\bar a_n,\bar b_n$ are independent Grassmann variables.
A basis is given by
\eqnl{
\psi_{n\pm}(x)= {1\ov\sqrt{V}}\;e^{i(p^\pm_{n},x)}\,e_\pm,
\mtxt{where} (p^\pm_n)_i={2\pi\ov L}(\ha+\al_i\pm\beta_i+n_i),}{dett30}
and the $e_\pm$ are the eigenvectors of $\gamfive$. Recall that $\al_i$
and $\beta_i$ represent the vectorial- and chiral twists \refs{t518}
respectively. The $\psi_{n+}$
and $\psi_{n-}$ must obey the ${\cal S}^+$ and ${\cal S}^-$
boundary conditions, respectively. These boundary conditions
fix the admissible momenta $p_n^\pm$ in \refs{dett30}. Since the
Dirac operator maps ${\cal S}^\pm$ into ${\cal S}^\mp$ the
$\chi_{n\pm}$ must then obey the same boundary conditions as the
$\psi_{n\mp}$. Thus $\chi_{n\pm}(x)$ is obtained from
$\psi_{n\pm}(x)$ by exchanging $p_n^+$ and $p_n^-$. It follows
then that
\eqnl{
i\hat\di\,\psi_{n\pm}=\lam_n^\pm\chi_{n\mp}}{t131a}
with
\eqngrl{
\lam_n^+&=& {2\pi\ov\tau_0L}
\Big[\bar\tau(\ha+a_1+\beta_1+n_1)-(\ha+a_0+\beta_0+n_0)\Big]}
{\lam_n^-&=& {2\pi\ov\tau_0L}
\Big[\tau(\ha+a_1-\beta_1+n_1)-(\ha+a_0-\beta_0+n_0)\Big].}{t31b}
Here we have introduced $a_\mu \equiv \al_\mu \!-\!g_{\sy 0}h_\mu\!-
\!\mu_\mu$. To continue we recast the infinite product for the
determinant in the form
\eqnl{
\prod_n^\infty\lam_n^+\lam_n^-=\prod_{{\vec{n}}\in Z^2} \Big(
{2\pi\ov L}\Big)^2\hat g^{\mu\nu}(\ha+c_\mu+n_\mu)(\ha+c_\nu+n_\nu),}{inf}
where $\hat g^{\mu\nu}$ is the inverse of the reference metric
\refs{t517} and
\eqnl{
c_\mu=a_\mu+i\hat\eta_\mu\,^\nu\beta_\nu,\mtxt{with} (\hat
\eta_\mu\,^\nu)=-{1\ov\tau_0}\pmatrix{\tau_1&-|\tau|^2\cr 1&
-\tau_1\cr}.}{t17a}
The logarithm of the product \refs{inf} can in turn be written as the
derivative at zero argument of a generalized zeta function. Indeed one
easily verifies that for
\eqnl{\zeta(s)\equiv\sum_n\big(\lam_n^+\lam_n^-\big)^{-s}}{det34}
we have (formally)
\eqnl{
\det(i\hat\di)\equiv \big(\prod_n\lam_n^+
\lam_n^-\big)_{reg}=\exp[-\zeta^\pr(s)]|_{s=0}.}{det341}
However $\zeta(s)$ is divergent for $s\leq 1$. These divergences can
be regularized as follows: We compute $\zeta(s)$ for $s> 1$ and
subsequently define its value for $s< 1$ by analytic
continuation.\hfill\break
Assume for the moment that $c_\mu$ is real or equivalently that
there are no chiral twists $\beta_\mu$ and chemical potential $\mu$.
Then $\zeta(s)$ has a well defined analytic continuation to $s\!<\!1$ via
a Poisson resummation \cite{Wipf91}. Indeed, writing  $\zeta(s)$ as a Mellin
transform
\eqnl{\zeta(s)=\frac{1}{\Gamma(s)}{\sum\limits_n}^\pr\int\,dt\,t^{s-1}
e^{-t\lam_n^+\lam_n^-},}{det5}
the generalized Poisson resummation formula
\eqnl{
\sum\limits_{Z}\exp[-\pi h^{\mu\nu}(n_\mu-a_\mu)(n_\nu-a_\nu)]=
\sqrt{h}\sum\limits_{Z}\exp[-\pi h_{\mu\nu}n^\mu n^\nu -2\pi i n^\mu
a_\mu],}{det6}
applied to the integrand in \refs{det5} yields after integration over $t$
\eqnl{
\zeta(s)=\frac{\Gamma(1-s)}{\Gamma(s)}\pi^{2s-1}
\sqrt{g}\,{\sum\limits_{Z}}^\pr(g_{\mu\nu}n^\mu
n^\nu)^{\frac{s-1}{2}}\exp[-2\pi i n^\mu (c_\mu+\ha)].} { det7}
The zero mode with $n_\mu\es 0$ is eliminated because for $s>1$ it
does not contribute. After this analytic continuation $\zeta(s)$ and
$\zeta^\pr(s)$ are now regular at $s\es 0$. More precisely $\zeta(0)= 0$ and
\eqngrl{\zeta^\pr(0)&=& \pi^{-1}
\sqrt{g}\,{\sum\limits_{Z}}^\pr(g_{\mu\nu}n^\mu
n^\nu)^{-\frac{1}{2}}\exp[-2\pi i n^\mu c_\mu]}
{&=&-\log\Big[ {1\ov |\eta(\tau)|^2}
\Theta\Big[{-c_1\atop c_0}\Big](0,\tau)\ \bar\Theta
\Big[{-c_1\atop c_0}\Big](0,\tau)\Big].}{t35}
Here we made use of $\det[c(i\hat\di)^2]\es \det[(i\hat\di)^2]$,
which follows from $\zeta(0)\es 0$.\pan
For complex $c_\mu$ the Poisson resummation
is not applicable and $\zeta^\pr (0)$ cannot be calculated
by direct means. To circumvent these difficulties we note that
the infinite sum \refs{det34} defining the $\zeta$-function
for $s\!>\!1$ is a mereomorphic function in $c$. Thus we
may first continue to $s\!<\!1$ for real $c_\mu$ and then
continue the result to complex values. Using the transformation
properties of theta functions the resulting determinant can
be written as
\eqnl{
\det(i\hat\di)= e^{2\pi(\sqrt{\hat
g}\hat g^{\mu\nu}\beta_\mu\beta_\nu-2i\beta_1a_0)}
 {1\ov |\eta(\tau)|^2}\Theta\Big[{-a_1+\beta_1\atop
a_0-\beta_0}\Big](0,\tau)\bar\Theta\Big[{-\bar a_1-\beta_1\atop
\bar a_0+\beta_0}\Big](0,\tau).}{t36}
This is the main result of this section.\hfill\break
It can be shown that this determinant is {\it gauge invariant},
i.e. invariant under $\al_\mu\rightarrow \al_\mu\!+\!1$, but
not invariant under chiral transformations, $\beta_\mu\rightarrow
\beta_\mu\!+\!1$, as expected. Furthermore, it transforms
covariantly under modular transformations $\tau\to\tau+1$ and
$\tau\to -1/\tau$. In other words, $\det i\hat\di$ is invariant
under modular transformations if at the same time the boundary
conditions are transformed accordingly. The exponential prefactor
is needed for modular covariance and is not present in the
literature \cite{chh4tfp88}. It correlates the two chiral sectors and will
have important consequences. In Appendix B we confirm \refs{t36} with
operator methods.\par
The last step
in the calculation of the fermionic generating functional is the
inclusion of the local contributions to the auxiliary- and
metric field, i.e. the dependence of the determinant on $\lam$,
$\phi$ and $\sigma$. For this we introduce
the one-parameter family of Dirac operators
\eqnl{\di_\tau=\frac{\hat g^{1/2}}{g_\tau^{1/2}}e^{\tau f^\dagger}\hat\di
e^{\tau f}.}{t539a}
We take the $\tau$-dependence of the metric as
$g_\tau\es e^{2\tau\sigma}\hat g$.
With $f$ as defined in
\refs{t28}, this family interpolates between $\hat\di$ and $\di$.
The determinant of the full
Dirac operator is then obtained by integrating the corresponding anomaly
equation \cite{chh4tp87}:
\eqnl{
&&{\det} i\di = {\det} (i\hat\di)
\exp\Big({S_L\ov 24\pi}+{g_{{\sy2 }}^2\ov 2\pi}\int\sqrt{\hat g}
\phi\hat\lap \phi\Big),}{t543}
where
\eqnl{
S_L=\int\sqrt{\hat g}\big[\hat\R\si -\si \hat\lap\si \big]}{t44}
is the {\it Liouville action.} In deriving this result we assumed
that $\int \sqrt{g}\lam\es 0$. This constraint on the zero-mode
of $\lam$ (and similarly of $\phi$) will be discussed below. Actually,
for our reference metric the Ricci scalar $\hat\R$ vanishes and the Liouville
action simplifies to $-\int\sqrt{\hat g}\si \hat\lap\si $.
However, the above formulae hold for arbitrary
reference metrics and arbitrary Riemannian surfaces.
Furthermore, as expected for a gauge-invariant regularization,
the function $\lam$ and thus the longitudinal part
of $B_\mu$ does not appear in the determinant. \par
To complete the calculation of the generating functional
we need to know the fermionic Green-functions $S$. Using the
scaling property of the Dirac operator, eq. \refs{t539a}, it is easy to
see that in an arbitrary background field $S$ is related to $\hat S$ by
\eqnn{
S(x,y)=e^{-f(x)}\;\hat S(x,y)\,e^{-f^\dagger (y)}.}
Together with the relation \refs{t543} and the explicit form
(\ref{t35},\ref{t36}) for
$\det i\hat\di$ this yields the fermionic generating functional
\eqngrl{
&&Z_F[\eta,\be]={1\ov |\eta(\tau)|^2}\Theta\Big[{-c_1\atop
c_0}\Big](0,\tau)\ \bar\Theta\Big[{-\bar c_1\atop
\bar c_0}\Big](0,\tau)}
{&&\quad e^{-\int\be(x) S(x,y)\eta(y)}\cdot\exp \Big(
{1\ov 24 \pi} S_L+{g_{{\sy 2}}^2\ov 2\pi}\int \sqrt{g}\phi\lap \phi\big]\Big).}
{t46}
By using the scaling properties of the Ricci-scalar and
Laplacian (see appendix A) the exponent can be written
in a manifest diffeomorphism-invariant way as
\eqnn{
-{1\ov 96\pi}\int\sqrt{g}\,\R{1\ov \lap}\R+{g_{{\sy 2}}^2\ov 2\pi}
\int \sqrt{g} \phi\lap \phi.}
Here we used that on the torus $\R$ integrates to zero.
On the sphere or higher
genus surfaces the last formula is modified.\pan
The Integration over the auxiliary fields then leads to the full generating
functional of the Thirring model. It contains all information about the
thermodynamic- and conformal properties. This is the subject of the next two
sections.

\subsection{Thermodynamics of the Thirring Model}
In this chapter we derive the grand canonical potential, equation
of state and ground state energy for the Thirring model.
For this we need to compute the partition function
\eqnl{
Z=\int d^2h\cd \phi\cd\lam \;Z_F[\eta\es\bar\eta\es 0]\;
e^{-S_B},}{t91}
where $Z_F$ is the fermionic generating functional \refs{t46}
and $S_B$ the bosonic action
\eqnl{
S_B=(2\pi)^2\sqrt{\hat g}\hat g^{\mu\nu}
h_\mu h_\nu-\int\sqrt{ g}\Big(\lam\lap\lam+\phi
\lap\phi\Big).}{t48}
As it stands the partition function is still ill-defined unless we
constrain the zero-modes artificially introduced in
the Hodge decomposition of $B_\mu$ in \refs{help}. The choice of the
constraints is restricted by the symmetries of the system.
In particular translation invariance (or rotation invariance on the sphere)
and covariance under modular transformations of the torus are symmetries
which me may want to preserve by the zero-mode constraint. The
constraint measure
\eqnl{
\int dh_0dh_1\cd\phi\cd\lam\delta(\bar\phi)\delta(\bar\lam)\cdots
\equiv \int dh_0dh_1\cd_\delta\phi\cd_\delta\lam\cdots,\qquad
\bar\phi\equiv {1\ov V}\int\sqrt{g}\phi}{52}
(and similarly for $\bar\lam$)
satisfies these requirements (The normalization by the volume in
the definition of $\bar\phi$ is needed such that
the constraints and hence the partition function are both
dimensionless). For example, one finds the dimensionless partition function
\eqnl{{\cal{N}}_0\equiv
\int \cd\phi\;\delta(\bar\phi)\,e^{(\phi,\lap\phi)}=
{\sqrt{V}\ov {\det}^{\pr\ha}(-\lap)}}{t53}
for free bosons, where the prime indicates the omission of the
zero-eigenvalue.\pan
\paragraph{Integration over the harmonics:} There is no restriction on the
harmonic parts of the auxiliary fields and the Gaussian integral yields
\eqnl{
\int\limits_{-\infty}^\infty d^2h \Theta\Big[{-c_1\atop c_0}\Big]
\bar\Theta\Big[{-\bar c_1\atop \bar c_0}\Big]
\exp[{-(2\pi)^2\sqrt{\hat g}\hat g^{\mu\nu}h_\mu h_\nu}]={\Theta\Big[
{u\atop w}\Big](\Lambda)\ov 4\pi \sqrt{1+g_{{\sy 0}}^2/2\pi}},}{td1}
where
\eqnn{
\Theta\Big[{u\atop w}\Big](\Lambda)=\sum\limits_{n\in Z^2}
\,e^{i\pi(n+u)\Lambda (n+u)+2\pi i (n+u)w}}
is the theta function with characteristics
\eqnl{
u=-\pmatrix{1\cr 1\cr}(\al_1+i\eta_1^{\;\nu}\beta_\nu)\mtxt{and}
w=\pmatrix{1\cr -1\cr}(\al_0+i\eta_0^{\;\nu}\beta_\nu-\mu_0)}{t93}
and covariance
\eqnl{
\Lambda=\pmatrix{\tau&0\cr 0&-\bar \tau\cr}
+i{\pi g_{{\sy 0}}^2\tau_0\ov 2\pi+g_{{\sy 0}}^2}\pmatrix{g_{{\sy 0}}^2&-
4\pi-g_{{\sy 0}}^2\cr
-4\pi-g_{{\sy 0}}^2&g_{{\sy 0}}^2\cr}.}{t94}
\paragraph{Integration over $\lam$ and $\phi$:}
The integral over $\lam$, subject to the $\delta$-constraint in \refs{52},
merely contributes one inverse square-root of the primed determinant
of $-2\lap$ to the partition function and so does the integration
over $\phi$. In fact,
to obtain the partition function of the Thirring model we divide $Z$ by the
corresponding partition functions ${\cal N}_0$ of the free
bosons, eq. \refs{t53}. Using \refs{td1} and \refs{t46} we obtain
\eqnl{{Z\ov {\cal N}_0}= {1\ov \vert\eta(\tau)\vert^2}
\sqrt{2\pi+ g_{{\sy 2}}^2\ov 2\pi+g_{{\sy 0}}^2}\;
\Theta\Big[{u\atop w}\Big](\Lambda)\,e^{(1/24\pi+g_{{\sy 3}}^2)S_L},}{t95}
where we have also used the scaling formula for the primed determinant of
$\lap$
\cite{chh4tbv89,WW}
\eqnl{
\log{{\det}^\pr(-a\lap)\ov {\det}^\pr(-\lap)}=
\log a\cdot\zeta(0)=\log a\cdot\big[{1\ov 4\pi}\int a_1-p\big],}{t58}
with $p$ being the number of zero modes of the operator. On
the torus $\int a_1\es 0$ and we find
\eqnn{
{\det}^\pr \big(-a\lap\big)={1\ov a}\;{\det}^\pr (-\lap),}
which produces the extra factor $\sqrt{2\pi+g_{{\sy 2}}}$.
In the Thirring model limit $ g_{{\sy 2}}\es g_{{\sy 0}}$ and the
square-root in \refs{t95} disappears.
\paragraph{Zero-temperature limit:}
To investigate the thermodynamics of the model we assume
space-time to be flat and that $\tau\es i\beta/L$. Then
\eqnn{
\Omega=-{1\ov \beta}\log {Z\ov{\cal N}_0} }
is the grand canonical potential. First we analyze the low
temperature limit of $\Omega$. For $\mu\es  0$ this yields the ground state
energy. We observe that for $\tau\es i\beta/L$  the
covariance matrix $\Lambda$ in \refs{t94} simplifies to
\eqnl{
i\pi\Lambda=-{\pi\beta\ov L}\Big[{\rm Id}+{g_{{\sy 0}}^2\ov 4\pi}{1\ov 2\pi+
g_{{\sy 0}}^2}
\pmatrix{g_{{\sy 0}}^2&-4\pi-g_{{\sy 0}}^2\cr-4\pi-g_{{\sy 0}}^2&g_{{\sy 0}}^2
\cr}\Big]}{t96}
and has eigenvalues
\eqnl{
\lam_1=-{\pi\beta\ov L}{2\pi+g_{{\sy 0}}^2\ov 2\pi}\mtxt{and}
\lam_2=-{\pi\beta\ov L}{2\pi\ov 2\pi+g_{{\sy 0}}^2}}{t97}
with corresponding eigenvectors
\eqnl{
v_1=(-1,1)\mtxt{and} v_2=(1,1).}{t98}
Also, the $\hat\eta$ tensor
(see \ref{t17a}) and $\mu_0$ (see \ref{t28}) in \refs{t93} simplify to
\eqnn{
\eta_\mu^{\;\;\nu}=\pmatrix{0&\beta/L\cr-L/\beta&0\cr}\mtxt{and}
\mu_0=-i{\beta\ov 2\pi}\mu.}
For $\beta\to\infty$ the saddle point approximation to the Gaussian sum
\refs{td1} defining the theta-function becomes exact and therefore using that
\eqnn{
\log\vert\eta(\tau)\vert^2\longrightarrow -{\pi\beta\ov 6 L}
\quad\mtxt{for}\beta\to\infty}
we find
\eqngrrl{
&&\Omega(\beta\to\infty)=-{\pi\ov 6L}-{4\pi\ov 2\pi+g_{{\sy 0}}^2}
{\pi\ov L}\big(\beta_1+{\mu L\ov 2\pi}\big)^2}
{&&\qquad\quad+{\pi\ov 2L}\min_{n\in Z^2}\Big[
{2\pi+g_{{\sy 0}}^2\ov 2\pi}\big\{n_2-n_1-{4\pi\ov 2\pi+g_{{\sy 0}}^2}(\beta_1
+{\mu L\ov 2\pi})\big\}^2}
{&&\qquad\qquad\qquad+{2\pi\ov 2\pi+g_{{\sy 0}}^2}\big\{
n_1+n_2-2\al_1\big\}^2\Big]}{t99}
for the zero-temperature grand potential of the un-gauged
model. Here the chemical potential and chiral twist enter
only through the combination $\beta_1\!+\!\mu L/2\pi$.
Let us now discuss the potential in the various {\it limiting cases}.\par
\subparagraph{i) No chiral twist, $\beta_1\es 0$, and
vanishing chemical potential:}
Then $\Omega(\beta\to\infty)$ coincides with the
{\it ground state energy}. The minimum in \refs{t99} is
attained for $n_1\es n_2\es [\ha\!+\!\al_1]$ and we find
\eqnl{
E_0(L,\al_1,\beta_1\es 0)=-{\pi\ov 6L}+{2\pi\ov L}{2\pi\ov 2\pi+g_{{\sy 0}}^2}
\big(\al_1-[\ha+\al_1]\big)^2.}{t100}
Only for anti-periodic boundary conditions, that is for $\al_1\es 0$,
does this {\it Casimir energy} coincide with the corresponding result for
free fermions. For $g_{{\sy 0}}^2\!\geq\!4\pi$ the Casimir force is always
attractive whereas for $g_{{\sy 0}}^2\!<\! 4\pi$ it can be attractive or
repulsive, depending on the value of $\al_1$.
The result \refs{t100} is in agreement with the literature
\cite{chh4tdv89}. For example, it coincides with De Vega's and Destri's result
if
we make the identification $\omega_{\sy DD}
\es 2 \pi \al_{\sy 1}$ and $1/\beta_{\sy DD} \es 1+ g_{{\sy 0}}^2/2\pi$
in formula (42) of that paper.
\pan
\subparagraph{ii) Small twists and chemical potential:}
For small $\beta_1$ and $\mu$ the minimum is assumed
for $n_i\es 0$ and the potential simplifies to
\eqnl{
\Omega(\beta\to\infty)=-{\pi\ov 6L}+{2\pi\ov L}{2\pi\ov 2\pi+g_{{\sy 0}}^2}
\al_1^2}{vir1}
and does not depend on the chemical potential. For vanishing $g_{{\sy 0}}$ the
minimum of \refs{t99} is attained for
\eqnn{
n_1=[\ha+\al_1-\beta_1-{\mu L\ov 2\pi}]\mtxt{and}
n_2=[\ha+\al_1+\beta_1+{\mu L\ov 2\pi}],}
where $[x]$ denotes the biggest integer which is smaller or
equal to $x$. This then leads to the following zero temperature potential
\eqngrrl{
\Omega=&-&{\pi\ov 6L} -{2\pi\ov L}(\beta_1+{\mu L\ov 2\pi})^2}
{&+& {\pi\ov L}\Big\{\al_1-\beta_1-{\mu L\ov 2\pi}
-[\ha+\al_1-\beta_1-{\mu L\ov 2\pi}]\big\}^2}
{&+&{\pi\ov L}\Big\{\al_1+\beta_1+{\mu L\ov 2\pi}
-[\ha+\al_1+\beta_1+{\mu L\ov 2\pi}] \big\}^2 .}{t101}
For $\mu\es \beta_1\es 0$ this reduces to the Casimir energy
for free fermions with left-right symmetric twists and agrees
with the results in \cite{chh4tkw93}.\pan
Note, however, that for $\beta_{\sy 1}\! \neq \! 0$
we disagree with \cite{chh4tdv89}. The difference is due to the second
term on the right in \refs{t99}.
Let us give two arguments in favour of our result: The discrepancy arises from
the prefactor appearing in the fermionic
determinant \refs{t36}. As discussed earlier this prefactor
implies the breakdown of holomorphic factorization, a property which
has been presupposed in \cite{chh4tdv89}. One can show that our
results can be reproduced by starting with massive fermions
and taking the limit $m\to 0$ (see appendix B).
\par \noindent
The second argument goes as follows: Suppose that $\beta_{\sy 1}\es
\al_1\es 0$. Then \refs{t101} simplifies to
\eqnl{
\Omega (\beta\to\infty)=-{\pi\ov 6L} -{2\pi\ov L}\Big({\mu L\ov 2\pi}\Big)^2
+ {2\pi\ov L}\Big({\mu L\ov 2\pi}-[\ha+{\mu L\ov 2\pi}] \Big)^2
.}{t102}
For massless fermions the Fermi energy is just $\mu$ and
at $T\es 0$ all electron states with energies less then $\mu$
and all positron states with energies less then $-\mu$ are filled.
The other states are empty. Since $d\Omega/d\mu$
is the expectation value of the electric charge in the
presence of $\mu$ we conclude that it must jump if $\mu$ crosses
an eigenvalue of the first quantized Dirac Hamiltonian $h$.
For vanishing twists the eigenvalues of $h$ are just $E_n\es(n-\ha)\pi/L $.
{}From \refs{t102} one sees by inspection that the electric charge
\eqnn{
\la Q\ra  ={d\Omega \ov d\mu}=2
\big[\ha+{\mu L\ov 2\pi}\big]=2n\quad\mtxt{for} E_n\leq\mu<E_{n+1}}
indeed jumps at these values of $\mu$. Further observe,
that in the thermodynamic limit $L\to\infty$ the density
\eqnn{
{\Omega\ov L}\rightarrow - {2\pi\ov 2\pi+g_{{\sy 0}}^2}{\mu^2\ov 2\pi}, }
reduces for $g_{{\sy 0}}\es 0$ to the standard result for free electrons.
\paragraph{Equation of state:}
We wish to derive the equation of state for finite $T$
in the infinite volume limit $L\to\infty$.
This may be achieved by interchanging the roles played by $L$ and
$\beta$. More precisely, using that
\eqnn{
\Theta\Big[{u\atop w}\Big](\Lambda)=\sqrt{\det(i\Lambda^{-1})}\;
e^{2\pi i w\cdot u}\;\Theta\Big[{-w\atop u}\Big] (i\Lambda^{-1})}
we find in analogy with the low temperature limit, that for
$L\to\infty$ the pressure is given by
\eqngrr{
\beta p&=&\lim_{L\to\infty}{1\ov L}\log {Z\ov {\cal N}_0}
={\pi\ov 6\beta}+{2\pi\ov \beta}{2\pi+g_{{\sy 0}}^2\ov 2\pi}\beta_0^2}
{&-&{\pi\ov 2\beta}\min_{n\in Z^2}\Big[
{2\pi+g_{{\sy 0}}^2\ov 2\pi}\big\{n_1+n_2+2\beta_0\big\}^2}
{&&\qquad\qquad +{2\pi\ov 2\pi+g_{{\sy 0}}^2}
\big\{n_2-n_1+2\al_0+2i{\beta\mu\ov 2\pi}\big\}^2\Big].}
Here the minimum of the real part has to be taken.
Again the minimization arises from the saddle point approximation
to the theta function which becomes exact when $L\to \infty$.
For small twists the minimum is assumed for $n_i\es 0$ and
then
\eqnn{
\beta p={\pi\ov 6\beta}-{2\pi\ov\beta}{2\pi\ov 2\pi+g_{{\sy 0}}^2}
\big(\al_0+i{\beta\mu\ov 2\pi}\big)^2}
becomes independent on the chiral twist $\beta_0$.
As we have interchanged the roles of the temporal and spatial
twists this is consistent with
the earlier result that for small twists $\Omega$ is independent of
$\beta _{\sy 1}$. In particular, for $\al_{\sy 0}\es 0$, we find
the following equation of state
\eqnl{
p(\beta,\mu,\al_0\es 0) ={\pi\ov 6\beta^2}+
{\mu^2\ov 2\pi}{2\pi\ov 2\pi+g_{{\sy 0}}^2}.}{vir2}
This result is consistent with the renormalization of the
electric charge which is conjugate to the chemical potential.
It shows that the thermodynamics of the Thirring model is
not just that of free fermions as has been claimed in
\cite{chh4ty87}. Indeed, the zero point pressure is multiplied by a
factor $2\pi/(2\pi+g_{{\sy 0}}^2)$.
This modification arises from the coupling of the current to the
harmonic fields. It is missed if only the local part of the
auxiliary field is considered, which is the
case if one quantizes the model in Minkowski space and then replaces
the $k_0$-integral in the Green functions by the Matsubara sum. This
remark should also be taken seriously in four dimensions!
Furthermore, we see that the 'pressure' $p$ is real only
for $\al_{\sy 0}\es 0$, which is consistent with the finite temperature
boundary conditions\footnote{This can also be observed in the
Hamiltonian formalism \cite{chh4tlt89}.}.

\subsection{Conformal structure }
In the first part of this section we derive the Kac-Moody and Virasoro
algebras of the model \refs{t1} without gauge-interaction and prepare the
ground for an extension, containing in particular the minimal models,
in the second part.\pan
Recall (\ref{t12a}) that for $A_\mu\es 0$
the theory reduces to a conformal field theory on flat Minkowski space-time.
To continue it is convenient to introduce adapted light cone
coordinates $x^\pm=x^0\pm x^1$
and the chiral components of the Dirac spinor $\psi_\pm=\ha
(1\pm\gamfive)\psi$. Then after substituting the classical equations of motion
\eqngrl{
T_{--}&=& -\ha(\pi_{\psi_+}\pa_-\psi_+-\pa_-\pi_{\psi_+}\psi_+)
+2(\pa_-\lam)^2+2(\pa_-\phi)^2}
{&&\qquad+i\pa_-(g_{{\sy 1}}\lam+ g_{{\sy 2}}\phi)
\pi_{\psi_+}\psi_+}{t14}
depends only on $x^-$ and is therefore the chiral Noether current.
Evaluating the Poisson bracket of
the symmetry generator $T_f=\int dx^-f(x^-)T_{--}$ with the different fields
yields the classical structure
\eqngrrl{
\delta_f\phi=f\pa_-\phi\qquad&;&\qquad \delta_f\lam=f\pa_-\lam}
{\delta_f\psi_+=f\pa_-\psi_++\ha\psi_+\pa_-f\quad &;&\quad
\delta_f\psi_+^{\sy\dagger}= \big(\delta_f\psi_+\big)^{\sy\dagger}}
{\delta_f j_{-} =f\pa_-j_{-}+j_{-}\pa_-f\quad&;&\quad
\delta_f T_{--}=f\pa_-T_{--}+2T_{--}\pa_-f.}{help1}
\paragraph{Short Distance Expansions:}
Let us now determine the quantum corrections to these
classical results. These are computed within the Euclidean functional
approach from the short-distance expansions of the relevant $n-$
point functions. We need not postulate Kac-Moody and Virasoro algebras
in advance as has been done in \cite{chh4tj64,chh4tfs89}.
These structures are derive here.
When comparing the classical with the quantum results one should
keep in mind that the roles of $\psi_{\sy 0} ^{\sy \dagger }$
and $\psi_{\sy 1} ^{\sy \dagger }$ are interchanged
when one switches from Minkowski to Euclidean space-time.
In coordinates adapted to the holomorphic structure of the torus
\eqnn{
x=i\bar\tau x^{\sy 0}+i x^{\sy 1}, \mtxt{so that}
\pa_x = {1\ov  2\tau_0}(\pa_{x^{\sy 0}} -\tau\pa_{x^{\sy 1}}),}
the Dirac operator and the corresponding Greens function take the form
\eqnn{
i\fdi  =2i\pmatrix{0&\pa_x\cr \pa_{\bar x}&0\cr}\mtxt{and}
S(x^\al,y^\beta)={1\ov 2\pi i}\pmatrix{0&1/\xi\cr
1/\bar \xi &0\cr}+O(1),}
where $\xi\es x\!-\!y$, and the chiral components of the energy momentum
tensor and current are given by
\eqnn{
T_{xx}={\tau_0\ov  2i}(\tau T^{\sy 00}+T^{\sy 01})=
{\tau_0\ov 2i}{d\hat g_{\mu\nu}\ov d\bar \tau}T^{\mu\nu}\mtxt{and}
j_x={1\ov  2i} (\tau j^{\sy 0}-j^{\sy 1}). }
{}From the conformal Ward identities
\eqnl{\sum\limits_{i=1}^n\big\la O(x_1)\cdots \de O(x_i)\cdots
O(x_n)\big\ra=\frac{1}{i}\oint dz \big\la O(x_1)\cdots
O(x_n)T_{zz}\big\ra}{ward}
we obtain the central charges and  conformal weights directly from the
correlation functions.
However, because on the flat torus the expectation value of $T_{xx}$ is
constant, we need to compute at least the $3$-point function to read
off the conformal weights. As in the classical theory (see \refs{t11}) the
symmetric energy momentum tensor measures the change of the effective action
$\Gamma=\log Z$ under arbitrary variations of the metric. On the
torus there are two independent contributions.
One being due to variations of the modular parameter $\tau$
and its conjugate $\bar\tau$ which
depend implicitly on the metric. The other is due to the variations
of terms which depend explicitly on the metric. Since the chiral component
$T_{xx}$ is gotten by contracting $T^{\mu \nu}$ with
$ d\hat g_{\mu\nu}/d\bar\tau$ it follows that
\eqnn{
\la T_{xx}\ra  ={i\tau_0\ov  \sqrt {g(x^\al)}}
\Big ({1\ov  L^2}{\pa\ov  \pa\bar\tau }  + {d\hat g_{\mu\nu}\ov
d\bar\tau} {\delta\ov  \delta g_{\mu \nu}(x^\al)} \Big )\;\Gamma
\lbrack g,\tau,\bar\tau \rbrack\equiv \delta_x
\Gamma\lbrack g,\tau,\bar\tau \rbrack.}
When doing metric variations it is always understood
that we take the flat space-time limit afterwards.
The $\bar\tau$ variation is constant and may be
discarded in the short distance expansion. Thus to analyze the algebraic
structure we can work on any Riemann surface. This is not true for the
finite size effects, which are global properties. This aspect will be
analyzed in section $3.4.$\pan
For example, taking three metric variations of the curvature dependent part
of $\log Z$ with $Z$ from \refs{t95} we find the following
short distance expansions for the three point correlation function
\eqnn{
\la T_{uu}\; T_{vv}\; T_{zz}\ra  \sim\;-
{1\ov (2\pi)^3}{1\ov (u-v)^2 (u-z)^2 (v-z)^2}.}
Substituting this result into the Ward identity \refs{ward} we obtain the
{\it central charge} and the conformal weight of the energy
momentum tensor
\eqnl{
c=1 \mtxt{and} h_{T_{xx}}=\;2\;.}{t104}
Note that the the central charge as well as the conformal weight are
independent of the couplings $g_{{\sy 1}}$ and $ g_{{\sy 2}}$. \pan
The conformal weights of the fundamental fields are obtained by
computing the fermionic two point function with stress tensor insertion
\eqnn{
\la \psi_{\sy 0}(x)\;\psi_{\sy 1}^{\sy\dagger}(y)\;T_{zz}\ra  =
{1\ov Z}\delta_z\Big(Z\la \psi_{\sy 0}(x)\;\psi_{\sy 1}^{\sy\dagger}(y)
\ra  \Big).}
Since $Z\sim\exp[F(\R^2)]$, its metric variation vanishes
after the flat space-time limit has been taken. The variation of
$S_{ij}$ can be found in appendix A. This yields
\eqngrl{
h_{ \psi_0}&=&h_{ \psi_1^\dagger }=\ha+{1\ov  16 \pi}  g_{{\sy 1}} ^2
-{1\ov  16\pi}{2\pi  g_{{\sy 2}}^2\ov  2\pi +   g_{{\sy 2}}^2 }}
{\bar h_{\psi_0}&=&\bar h_{\psi_1^\dagger }={1\ov  16 \pi}  g_{{\sy 1}} ^2
-{1\ov  16 \pi} {2 \pi   g_{{\sy 2}} ^2 \ov  2 \pi +   g_{{\sy 2}}^2 }
\;.} {t507}
Thus we have reproduced the classical results supplemented
by additional $ g_{{\sy 1}} $ and $  g_{{\sy 2}} $ dependent quantum
corrections.
In the Thirring model limit $ g_{{\sy 2}} \es   g_{{\sy 1}} \es g$,
these terms add up to give the known anomalous dimension
appearing in the Thirring model \cite{chh4tfs89}. Furthermore, from
\refs{t507} we may derive a condition on the couplings $g_{{\sy 1}},
g_{{\sy 2}}$ if we insist on unitarity, i.e. on $h\geq 0$. We find
\eqnl{g_{{\sy 1}}^2\geq \frac{2\pi  g_{{\sy 2}}^2}{2\pi + g_{{\sy 2}}^2}.}{tz3}
In particular for $g_{{\sy 1}}\geq \sqrt{2\pi}$ the conformal weights are
positive for any real $ g_{{\sy 2}}$.\pan
Next we determine the {\it Kac-Moody algebra} of the
$U(1)$ currents. To derive the correlation functions with current
insertions we couple the fermions to an external vector field, that is
consider the 'gauged' model without Maxwell term. For example,
\eqnn{
<j^{\mu}(x^\al)\;j^\nu(y^\beta)>\;=
{1\ov e^2\sqrt{g(x^\al)g(y^\beta )}}\;{\delta^2 \Gamma \lbrack g,A\rbrack
\ov \delta A_\nu(x^\al)\delta A_\mu(y^\beta)}\vert_{A=0}. }
The effective action with external vector field is then obtained by shifting
the auxiliary fields in \refs{t28} as
\eqnl{g_{{\sy 2}}\phi\rightarrow g_{{\sy 2}}\phi+e\varphi\mtxt{,}
g_{{\sy 1}}\lam \rightarrow g_{{\sy 1}}\lam +e\alpha,}{kmn1}
where $A_\mu\es \eta_\mu^{\;\;\nu}\pa_\nu\varphi +\pa_\mu\alpha$ and we have
neglected the harmonic contribution to the external vector field, because it
does not contribute to the short distance expansion. The resulting effective
action does not depend on $\alpha$ due to gauge-invariance. To relate the
variation w.r.t. $A_\mu$ to that w.r.t. $\varphi$ we use
\eqnn{
\pa_\mu \phi=\eta_\mu^{\;\,\nu}A_\nu^{\sy T},\mtxt{where}
A_\mu^{\sy T}=A_\mu-\nabla_\mu{1\ov\lap}\nabla^\nu A_\nu
}
is the transverse part of $A_\mu$. We obtain the following short distance
expansion
\eqnn{
\la j_x\;j_y \ra  \; \sim -{1\ov 2\pi}{1\ov 2\pi + g_{{\sy 2}}^2}
{1\ov (x-y)^2}\;.
}
We read off the value $k$ of the {\it central extension} in the
$ U(1)$-Kac-Moody algebra
\eqnl{
k ={2\pi \ov 2\pi +   g_{{\sy 2}} ^2 }.}{t105}
The precise $g_{{\sy 2}}$-dependence of $k$ (which can of course be
rescaled to unity by an appropriate redefinition of the current) is
related to a finite renormalization of the electric charge in the gauged
Thirring-model which we will discuss in section $4$.\pan
Finally, from
\eqnn{
\la j_x \; j_y \; T_{zz}\ra\;\sim-{1\ov 4\pi^2}{1\ov 2\pi + g_{{\sy 2}}^2}
{1\ov (x-z)^2(y-z)^2} }
we obtain $h_j=1$.\par
To see how the left and right Kac Moody currents act on the fermionic fields
we notice that after the integration over the auxiliary fields
the $A$-dependence of the fermionic Green function factorizes as
\eqnn{
\la \psi_0(x)\psi_1^{\sy \dagger}(y)\ra  _{\sy A}=
e^{\ha m_\gam \int   \varphi\lap   \varphi}\cdot e^{-eg(x)}\;\la \psi_0(x)
\psi_1^{\sy \dagger}(y)\ra  _{\sy A=0}\;e^{-eg^\dagger (y)},}
where $g(x)=-i\al(x)+\gam_5\beta \varphi(x)$,
$\beta=2\pi/ (2\pi+ g_{{\sy 2}}^2)$ and $m_\gam$ is the induced
'photon'-mass (see\refs{t60}). Variation w.r.t. the $A-$ field yields,
after some algebraic manipulations, the $U(1)$ charges
\eqnl{
q_{{\psi_0}}=\ha\big(1+{2\pi\ov 2 \pi+  g_{{\sy 2}}^2}\big)\mtxt{and}
\bar q_{{\psi_0}}=\ha\big(1-{2\pi\ov 2\pi+  g_{{\sy 2}}^2}\big).}{t110}
We have used the convention where the electric
charge $q\!+\!\bar q$ is unity. In the Thirring model limit we
can compare \refs{t110} with the results obtained in \cite{chh4tfs89}.
For that we need to rescale the currents such that the central
extension \refs{t105} of the Kac-Moody algebra becomes unity $
j_z \to \sqrt {1+  g_{{\sy 2}}^2/2\pi} \; j_z \;$ .
It is then easy to see that we agree with Furlan et al. \cite{chh4tfs89}
if we make the identification $\bar g_{\sy {Fu}}\es
g_{\sy 2}^2/4\pi\sqrt{1+g_{{\sy 2}}^2/2\pi}$.\par
\paragraph{Non-Minimal Coupling:}
In section $3.1$ we have analyzed the fermionic determinant in the presence
of twisted boundary conditions. One may ask what happens if we introduce a
local twist instead, that is
\eqnl{\psi(x)\rightarrow \psi(x)\mtxt{;}\psi(x)^\dagger\rightarrow
\psi(x)^\dagger e^{\al\lam(x)},}{twist1}
which should be interpreted as a modification of the charge neutrality
condition. The computation of the fermionic determinant in the presence of
such twists is similar to that for a Weyl rescaling of the background metric
(\ref{t539a}-\ref{t543}). Integrating the corresponding anomaly equation we
find
\eqnl{\log\frac{\det(i\di_\al)}{\det(i\di_0)}\propto \al\int\R\lam +
O((\al\lam)^2).}{twist2}
We will come back to the relation between the above determinant and charges at
infinity at the end of this section. For the moment we use the analogy
merely as a motivation to study the extension of the Thirring model obtained
by coupling the $\lam$-field non-minimally to the background geometry. That
is we consider the model \refs{t1} again without gauge-interaction but with
an extra coupling
\eqnn{g_{{\sy 3}}\int\R\lam.}
Then $T_{--}$ in \refs{t14} is modified,
\eqnn{
T_{--}\longrightarrow \bar T_{--}=T_{--}+3g_{\sy 3}\pa_-^2\lam.}
The corresponding modification of the classical conformal transformations
\refs{help1} generated by the modified generator
$\bar T_f=\int dx^- f(x^-)\bar T_{--}$ are
\eqngrl{
\bar\delta_f\phi=\delta_f\phi \qquad &,&\qquad
\bar\delta_f\lam=\delta_f\lam -{g_{{\sy 3}}\ov 2}\pa_-f}
{\bar\delta_f\psi_+=\delta_f\psi_+-{i\ov 2}g_{{\sy 1}}g_{{\sy 3}}\psi_+\pa_-f
\quad&,&\quad\bar\delta_f\psi_+^{\sy\dagger}=\delta_f\psi_+^{\sy\dagger}
+{i\ov 2} g_{{\sy 1}}g_{{\sy 3}}\psi_+^{\sy\dagger}\pa_-f.}{help2}
Whereas $\phi$ and $\psi_+$ remain primary fields, $\lam$ does not. This
is in fact needed for consistency. Indeed, since $\psi$
is not a scalar under conformal transformations generated
by $\bar T_f$, the term $\sim \int \psi^\dagger \di\psi$ in the action
is only conformally invariant if $\lam$ transforms inhomogeneously
like a spin connection.\par
It may be surprising that the new symmetry transformations depend on the
coupling constant $g_{{\sy 3}}$ which is not present in the flat space time
Lagrangian. However, the same happens for example in $4$ dimensions,
if one couples a scalar field conformally, that is non-minimally, to
gravity. Although the Lagrangian for the minimally and conformally
coupled particles are the same on Minkowski space-time, their
energy momentum tensors are not. The same happens for the conformally
invariant non abelian Toda theories which admit several energy
momentum tensors and hence several conformal structures \cite{chh4trw90}.\pan
The current still transforms as a primary with weight $1$, but
the energy momentum tensor acquires a classical central charge
\eqnl{
\bar\delta_f \bar T_{--}=f\pa_-\bar T_{--}+2\bar T_{--}\pa_-f
-g_{{\sy 3}}^2\pa_-^3f.}{t16b}
The corresponding commutators in the quantized theory with
non-minimal coupling to gravity are calculated as explained for
the minimally-coupled model. One finds that the quantum corrections
to \refs{help2} are identical to those of the minimally coupled model and
thus are $g_{\sy 3}\neq 0$-independent.\pan
To summarize, we have obtained the following Virasoro $\times$ Kac-Moody
structure:\pan
Central charge:
\eqnl{c=1+24 g_{{\sy 3}}^2 \pi \mtxt{and} h_{T_{xx}}=\;2}{help3}
Kac-Moody level and charges:
\eqngr{k={2\pi\ov 2\pi+g_{{\sy 2}}^2}\qquad&;&\qquad h_j=1}
{q_{{\psi_0}}=\ha\big(1+{2\pi\ov 2 \pi+  g_{{\sy 2}}^2}\big)\quad &;&\quad
\bar q_{{\psi_0}}=\ha\big(1-{2\pi\ov 2\pi+  g_{{\sy 2}}^2}\big)}
Conformal weights:
\eqngrl{
h_{ \psi_0}&=&\;\ha+{1\ov  16 \pi}  g_{{\sy 1}} ^2
-{1\ov  16\pi}{2\pi  g_{{\sy 2}}^2\ov  2\pi +   g_{{\sy 2}}^2 }
-{i g_{{\sy 1}}   g_{{\sy 3}} \ov  2}
=\big(h_{ \psi_1^\dagger }\big)^\dagger}
{\bar h_{\psi_0}&=&\;{1\ov  16 \pi}  g_{{\sy 1}} ^2
-{1\ov  16 \pi} {2 \pi   g_{{\sy 2}} ^2 \ov  2 \pi +   g_{{\sy 2}}^2 }
-{i g_{{\sy 1}}   g_{{\sy 3}} \ov  2}=\big(\bar h_{\psi_1^\dagger}\big)^\dagger
.} {t107}
Here some comment about unitarity is in order. It can be
shown that with respect to the standard scalar product \cite{OS}
reflection-positivity holds for any real $g_{{\sy 3}}$ \cite{thesis}. However
with respect to this
inner product the
Virasoro generators are not selfadjoint. Choosing an alternative scalar product
\cite{chh4tg94} for which they are selfadjoint, positivity does not hold in
general
for $g_{{\sy 3}}\neq 0$. We give a more detailed discussion about unitary
subspaces
in section $3.5$.

\subsection{Finite size effects}
When quantizing a conformal field theory on a space-time
with finite volume one introduces a length scale.
The presence of this length scale in turn breaks the conformal
invariance and gives rise to finite size effects. It has
been conjectured \cite{chh4tc88} that the finite size effects on a
Riemann surface
are proportional to the central charge. For example,
when one stretches space time, $x^\al\to a x^\al$, then
the change of the effective action is proportional to $c$:
\eqnl{
\Gamma_{ax}-\Gamma_x=-{c\ov 6}\log a\cdot\chi,}{t111}
where $\chi$ is the Euler number of the Euclidean space time.
In \cite{chh4tdw92} this conjecture has been proven for
a wide class of conformal field theories on spaces with boundaries.
The only important assumption has been that the regularization
respects general covariance. In this subsection we shall see
that the equivalence does hold only for a particular zero-mode treatment,
which differs from \refs{52}.\par
The only global conformal transformations on the
torus are translations which do not give rise to finite size
effects. Also, the Euler number vanishes and according to
\refs{t111} the finite size effects are insensitive to the value of $c$.
For that reason we quantize the un-gauged model \refs{t1} on the
sphere where the global conformal group is the Moebius group.\par
An effective method to compute finite size effects has been
developed in \cite{chh4tdw92}. It is based on the following
observation: Any conformal transformation $z\to w(z)$ is a composition
of a diffeomorphism (defined by the same $w$) and a compensating
Weyl transformation $g_{\mu\nu}\to e^{2\si }g_{\mu\nu}$
with
\eqnn{
e^{2\si }={dw(z)\ov dz}{d\bar w(\bar z)\ov d\bar z},\qquad
z=x^0+ix^1.}
Therefore, choosing a diffeomorphism invariant regularization
one has
\eqnn{
0=\delta \Gamma_{Diff}=\delta\Gamma_{Conf}-\delta\Gamma_{Weyl}.}
The change of the effective action under Weyl rescaling is
\eqnn{
\delta\Gamma_{Weyl}=-\log {\int{\cal D}(\lam\phi)\det (i\di_g )
\exp (-S_B [g])\ov
\int{\cal D}(\lam\phi)\det (i{\di}_{\hat g})\exp (-S_B[\hat g])}\;, }
where $S_B$ is the bosonic action \refs{t48}. Since on the sphere there are no
harmonic
vector fields the term $\sim h^2$ in $S_B$ is not present. Imposing
the conditions \refs{52} we obtain
\eqnl{
\delta\Gamma_{Weyl}=\log {\hat V\ov V}-{S_L\ov 24\pi}
+{g_{{\sy 3}}^2\ov 4}\int \R{1\ov \lap}(\R-\frac{8\pi}{V})+
\log{\det^\pr \triangle\ov \det^\pr \hat\triangle}.} {t112}
To evaluate \refs{t112} one introduces the $1$-parametric
family of Laplacians
\eqnn{
\lap_\tau=e^{-2\tau\si }\hat\lap}
interpolating between $\hat\lap$ and $\lap$. Integrating the corresponding
anomaly equation \cite{chh4tbv89} we end up with
\eqnl{
\delta\Gamma={ g_{{\sy 3}}^2\ov 4}\int\sqrt{g}\R\ilap \Big(\R-{8\pi\ov V}\Big)
-{3\ov 24\pi }\int\sqrt{\hat g}\hat\R\si+{3\ov 24\pi}\int\sqrt{\hat g}
\si\hat\lap\si.}{t114}
Consider now a dilatation $w(z)\es a z$. Then, the
conformal angle is constant, $\si\es\log a$, and $(\R-8\pi/V)\es 0$. Then
the first term in \refs{t114} vanishes and the finite
size effect does not depend on $  g_{{\sy 3}} ^2$. It is
given by
\eqnn{
\delta\Gamma=-{3\ov 24\pi}\log a\int\sqrt{\hat g}\hat\R=-\log a}
and does not agree with \refs{t111} since $c$ in \refs{help3}
depends on $g_{{\sy 3}}$. On other
Riemannian surfaces one would find the same result. Note
that the finite size scaling comes from the middle term
$\sim \log a\int \sqrt{\hat g}\hat R$ in \refs{t114} which is topological in
nature, while the short-distance behaviour of the energy-momentum correlators
is controlled by the remaining two terms in \refs{t114} which are insensitive
to the topology. In that sense finite size scaling and the central
charge are complementary. There is a way to match the two results by adding
the term
\eqnn{
-{  g_{{\sy 3}}^2\ov 4}\int \sqrt{g}\R\lap\R}
to the effective action. With this new effective action the short
distance expansion of the energy-momentum correlators does not depend
on $g_{{\sy 3}}$ any more and the corresponding central charge equals that
obtained from the finite size scaling. However such a term would
correspond to a non-local counter term to be added to the regularized action.
\subsection{Charge neutrality and unitary subspaces}
In this subsection we show how the equivalence between the central charge
and finite size scaling
can be restored, provided the partition function is replaced by an average
over un-normalized expectation values of charges at infinity. In fact it
turns out that the $g_{{\sy 3}}\int\R\lam$-term, ie. the non-minimal coupling
to gravity, itself can be given the interpretation of a charge at infinity if
the zero-mode constraints \refs{52} is replaced by a non-translation invariant
sum over charges at infinity.\par
The hint comes from inspecting the fermionic weights \refs{t107}, which shows
that $\psi(x)$ and $\psi_{g_{{\sy 3}}}(x)\equiv e^{-8\pi g_{{\sy 3}}\lam(x)}
\psi(x)$
have the same conformal weights. We can therefore consistently put a
charge at infinity with a corresponding modification of the charge
neutrality condition. The non-vanishing two-point function is
now $\la\psi_{g_{{\sy 3}}}(x)^\dagger\psi(x)\ra$. It's coincidence limit
$j_{g_{{\sy 3}}}$ is again a primary field with conformal weight
$h_j\es 1$.\pan
On the other hand, including a charge at infinity into the definition of the
partition function we have
\eqngrl{Z_{g_{{\sy 3}}}&=&\frac{1}{N_0}\int \cd_\de \phi\cd_\de\lam  \;
Z_F[\eta\es\bar\eta\es 0]\;
e^{-S_B}\; :e^{8\pi g_{{\sy 3}}\lam (\xi_0)}:}
{&&=Z_0\;\exp[16\pi^2 g_{{\sy 3}}^2G_0(\xi_0,\xi_0)]\qquad\mtxt{(recall
that}\cd_\de\phi
=\delta(\bar\phi)\cd\phi).}{t591}
To continue we need to determine the coincidence limit of the scalar Greens
function $G_0(x,y)$, i.e. to regularize the composite operator
$\exp(\al\lam)$ appearing in \refs{t591}. The normal ordering
prescription
\eqnl{
:e^{\al\lam(x)}:= {e^{\al\lam(x)}\ov\la e^{\al\lam(x)}\ra}.}{t76}
works well on the whole plane \cite{chh4tc75,chh4tsg86}. On curved
space we must be more careful when renormalizing this operator.
The required wave function renormalization is not unique
but it is very much restricted by the following
requirements: First we take as reference system
(the denominator in \refs{t76}) one with a minimal number
of dynamical degrees of freedom since we do not want to
loose information by our regularization. Second, the renormalized
operator should have a well-defined infinite volume limit. Finally, the
regularization should respect general covariance.
These requirements
then force us to take as reference system the infinite plane
with metric $g_{\mu\nu}$. The flat metric $\delta_{\mu\nu}$
is not permitted since it leads to a ill-defined expression
for $\la\exp(\al\lam)\ra$. With this choice
the normal ordering in \refs{t76} is equivalent to replacing
the massless Green function in \refs{t591} by
\eqnl{
G_0^{reg}(x,y):= G_0(x,y)+{1\ov 4\pi}\log\big[\mu^2 s^2(x,y)\big].}
{t77}
Here $s(x,y)$ denotes the geodesic distance between $x$ and $y$.
The occurrence of the arbitrary mass scale $\mu$ comes from the
ambiguities in the required ultra-violet regularization. On
the $2$-sphere with constant Ricci scalar $\R$ we have
\eqn{G_0^{reg}(x,x)\es-\frac{1}{4\pi}[\log[\frac{\R}{8\mu^2}]+1].}
The expectation value $\la :e^{8\pi g_{{\sy 3}}\lam(\xi_0)}:\ra$ then
transforms under a constant rescaling $z\rightarrow az$ as
\eqnl{\la :e^{8\pi g_{{\sy 3}}\lam(\xi_0)}:\ra\rightarrow
\la :e^{8\pi g_{{\sy 3}}\lam(\xi_0)}:\ra\exp[8\pi g_{{\sy 3}}^2\log(a)],}{t594}
and therefore gives an extra contribution
\eqn{\de\Gamma_{g_{{\sy 3}}}=-\frac{24\pi g_{{\sy 3}}^2}{6}\log(a)\chi,}
to the finite size scaling of the effective action. Adding this
contribution to \refs{t114} above we see that this is precisely the
piece needed to restore equivalence with the central charge for any
real or imaginary $g_{{\sy 3}}$.\pan
More generally we can define the functional integral as an average over all
possible
charges at infinity: assume $g_{{\sy 3}}$ imaginary. The
(un-normalized) expectation values are then given by
\eqnl{\Big\la\prod_{i=1}^n O_{\al_i}(x_i)\Big\ra\equiv\frac{1}{Z}\int\cd_\de
\phi\cd\lam\Big[\frac{1}{\sqrt{2\pi}}\int
d\;k\;:e^{ik\lam(\xi_0)}:\Big]\;\prod_{i=1}^n  O_{\al_i}(x_i)\;
e^{-S_B}.}{t555}
Here $\al_i$ denotes the $U(1)$-charge of the operator $O_i$.
In particular the partition function on $S^2$ is
\eqnn{Z=\frac{1}{N_0}\int\cd_\de \phi d\lam_0\cd\lam^\prime \Big[\frac{1}
{\sqrt{2\pi}}\int d\;k\;:e^{ik\lam(\xi_0)}:\Big]\;:e^{-8\pi g_{{\sy 3}}
\lam(\xi_0)}:\; \;e^{-S_B[\lam ^\prime]},}
where $\lam_0$ is the zero mode and $\lam^\prime$ the excited modes of
$\lam (x)$. The middle term in the above integrand is the zero-mode
part of $S_B$. The zero-mode integration yields a delta function $\de(k+i8\pi
g_{{\sy 3}})$ and thus the $g_{{\sy 3}}\int\R\lam$-term itself
acquires the interpretation of a charge
at infinity, due to the presence of the zero mode. The 'extra' charges
$e^{ik\lam(\xi_0)}$ assure the charge neutrality of the partition function.
For the general n-point function \refs{t555} the $\lam_0$- integration yields
\eqn{
\de(k+8\pi i g_{{\sy 3}}+\sum\limits_{i=1}^{n}\al_i),}
where the sum of the $U(1)$-charges of the operators in
\refs{t555} enters. In particular, for neutral states, for which ($\sum
\al_i+i8\pi
g_{{\sy 3}}=0$), $k$ must be zero and no extra charge at infinity
is present.\pan
Finally, using
\eqn{\frac{1}{\sqrt{2\pi}}\int d\;k\;e^{ik\lam(\xi_0)}=\de(\lam(\xi_0)),}
the averaging over all possible charges can also be written as
\eqnl{\cd\lam\de(\lam(\xi_0)).}{t666}
It is easy to verify that if the action has translation invariance in the
target space, then the constraints \refs{t666} and \refs{52}
are equivalent and the correlation functions do not depend on
the chosen base-point $\xi_0$. However, in the present
case \refs{t666} clearly breaks translation invariance (or rotation
invariance on $S^2$) and the zero-modes constraints are inequivalent.
Although we have assumed an imaginary $g_{{\sy 3}}$,
our results apply for any $g_{{\sy 3}}$.
For particular values we recover the
(unitary) minimal models, provided screening charges \cite{Banks} are included
for the
$n$-point function with $n>2$.
In particular for $g_{{\sy 3}} = 1/\sqrt{48\pi}$ and $g_{{\sy 1}}\es
g_{{\sy 2}}\es 0$ we obtain the Ising model with
$h_\psi\es h_{\psi^\dagger}\es \ha$.
\section {Gauged Thirring-like Models}
In this section we extend the model by gauging the global $U(1)$-symmetry.
Contrary to what one might think, many aspects of the gauged model are
actually simpler as compared to the ungauged model.
In particular the thermodynamical properties are independent of external
conditions like chemical potentials and twisted
boundary conditions. The reason is that the model is closely related
to the Schwinger model, for which the spectrum consist solely of a
neutral, massive particle. On the other hand, the gauge interaction
complicates the analysis, because the $U(1)$- bundle over the torus allows
for gauge field configurations with winding number,
so called instantons. These, in turn, imply fermionic zero-modes which
trigger a chiral symmetry breaking
and therefore a non-vanishing condensate. This is the subject of the second
part of this section. In
the first part we discuss the partition function to
which only topologically trivial configurations contribute.\par
To see how the fermionic generating functional \refs{t46} is modified,
we decompose a general gauge potential on a torus as
\eqnl{
A_\mu=A^I_\mu+{2\pi\ov L}t_\mu+\pa_\mu\al-\eta_{\mu\nu}
\pa^\nu  \varphi,}{t520}
where the last $3$ terms correspond (as for the auxiliary field $B_\mu$) to
the Hodge decomposition
of the single valued part of $A$ in a given topological
sector, that is the harmonic-, exact- and co-exact pieces.
The role of the toron fields $t_\mu$ has recently been
emphasized within the canonical approach \cite{chh4tln93}. In the
Hamiltonian formulation they are quantum mechanical degrees
of freedom which are needed for an understanding of the
infrared sector in gauge theories. Also, in \cite{chh4ts93} it has been
argued that the $Z_N$-phases of hot pure Yang-Mills
theories \cite{chh4ts79} should correspond to the same physical state
if the toron fields are taken into account.
The first term in \refs{t520} is an {\it instanton potential}
which gives rise to a non-vanishing quantized flux. As noted above
configurations with non-vanishing flux do not contribute to the partition
function due to the associated fermionic zero modes. We can therefore assume
$A^I_\mu\es 0$ for the moment.
The fermionic generating functional is obtained from \refs{t36} by simply
shifting
\eqnn{
g_{{\sy 0}}h_\mu\to et_\mu+ g_{{\sy 0}}h_\mu\es H_\mu\mtxt{,}
\F \to e\al+\F \es F\mtxt{and}\G\to\G+e\varphi \es G,}
which leads to
\eqngrl{
&&Z_F[\eta,\bar\eta]=e^{2\pi(\sqrt{\hat
g}\hat g^{\mu\nu}\beta_\mu\beta_\nu-2i\beta_1a_0)} {1\ov |\eta(\tau)|^2}
\Theta\Big[{-a_1+\beta_1\atop
a_0-\beta_0}\Big](0,\tau)\bar\Theta\Big[{-\bar a_1-\beta_1\atop
\bar a_0+\beta_0}\Big](0,\tau)}{&&\quad e^{-\int\be(x) S(x,y)\eta(y)}\cdot
\exp \Big(
{1\ov 24 \pi} S_L+{1\ov 2\pi}\int \sqrt{g}G\lap G\big]\Big),}{t636}
with $a_\mu\es\al_\mu-H_\mu-\mu_\mu$.\par
To compute the {\it partition function}
we must switch off the sources $\eta$ and $\bar\eta$ in
\refs{t636} so that
\eqnl{
Z_0 = J\int d^2td^2h\cd    \varphi\cd\phi\cd\lam
\;Z_F[0,0]\; e^{-S_B},}{t54}
where now
\eqngrl{
S_B&=&(2\pi)^2\sqrt{\hat g}\hat g^{\mu\nu}
h_\mu h_\nu}
{&+&\int\sqrt{ g}\Big(\ha   \varphi\lap^2   \varphi-\lam\lap\lam-\phi
\lap\phi-g_{{\sy 3}}R\lam\Big).}{t648}
Note that we have kept the non-minimal coupling of the $\lam$-field to
gravity as in section $3.3$.
Since $S_B$ and the fermionic determinants are both gauge
invariant and thus independent of the pure gauge mode
$\al$ in \refs{t520}, it is natural to change variables from
$A_\mu$ to $(   \varphi,\al,t_\mu)$. This transformation is
one to one, provided
\eqnl{
\int\sqrt{g}   \varphi=\int\sqrt{g}\al=0\mtxt{and}et_\mu\in [0,1].}{t49}
In contrast to the auxiliary harmonic fields $h_\mu$ in \refs{help},
the toron fields $et_\mu$ and $et_\mu+n_\mu$ with
integer $n_\mu$ are to be identified, due to gauge invariance
\cite{chh4tsw92}. The measures are related as
\eqnl{
\cd A_\mu=J\sum_k\;dt_0dt_1\cd   \varphi\cd\al,\mtxt{where}
J=(2\pi)^2{\det}^\pr (-\lap).}{t50}
In
expectation values of gauge invariant and thus $\al$-independent
operators the $\al$-integration cancels against the normalization.
This simply expresses the fact that in $QED$ the ghosts
decouple in the Lorentz gauge.\pan
As we shall see shortly it is advantageous to integrate first over the toron
fields. By using the series representation of the theta functions
one computes
\eqnl{\int\limits_0^1 d^2(et)\Theta\Big[{-c_1\atop
c_0}\Big](0,\tau)\ \bar\Theta\Big[{-c_1\atop
c_0}\Big](0,\tau)={1\ov \sqrt{2\tau_0}}.}{t55}
Since the result appears always together with the $\eta$-function
factor in \refs{t46} it is convenient to introduce
\eqnn{
\kappa :={1\ov \sqrt{2\tau_0}}{1\ov
\vert\eta (\tau)\vert^2}}
in the following expressions. The result \refs{t55} does not
depend on the $h$-field and hence the $h$-integration in
\refs{t54} becomes Gaussian and yields a factor $1/4\pi$ so that
\eqnl{
Z_0=\pi\kappa\,{\det}^\pr(-\lap)\;e^{S_L/24\pi}\int
\cd_\delta (   \varphi\phi\lam)\;
e^{{1\ov 2\pi}\int \sqrt{g}G\lap G-S_B[h=0]},}{t56}
where we inserted
the explicit expression \refs{t50} for the Jacobian.
Now we see why we did well integrating over the toron fields first. It has
washed out the
dependence on the boundary conditions and chemical potential in \refs{t55}.
\pan
The integral over $\lam$, subject to the condition in \refs{52}, decouples
completely apart from the non-minimal coupling to gravity which
modifies the Liouville factor and yields one inverse square-root
of the determinant of $-2\lap$ in \refs{t56}. Thus
\eqngrl{
Z_0&=&\kappa\pi\sqrt{2V{\det}^\pr(-\lap)}\;e^{(g_{{\sy 3}}^2+1/24\pi)S_L}}
{&&\qquad\quad\cdot\int\cd_\delta (   \varphi\phi)\;e^{{1\ov 2\pi}
\int \sqrt{g}G\lap G-S_B[h=\lam=0]},}{t59}
where we have used \refs{t58}.
The $\phi${\it -integration} in contrast, leads to a finite
renormalization of the
dynamically generated 'photon' mass
\eqngrl{
&&Z_0={2\sqrt{\pi}\kappa e V\ov m_\gam}\;
e^{(g_{{\sy 3}}^2+1/24\pi)S_L}
\int\cd   \varphi e^{-\ha\int\sqrt{g}   \varphi\big(
\lap^2-m_ \gamma^2\lap\big)   \varphi},}{&&\mtxt{where}
m_ \gamma^2={e^2\ov \pi}{2\pi\ov 2\pi+ g_{{\sy 2}}^2}}{t60}
plays the same role as the $\eta^\pr$-mass in $QCD$ \cite{chh4gs93}.
The determinant obtained from the $\varphi-$ integration factorizes as
\eqnn{
{\det}^\pr (\lap^2-m_ \gamma^2\lap)={\det}^\pr(-\lap)\cdot
{\det}^\pr (-\lap+m_ \gamma^2).}
This factorization property is not obvious since all
determinants must be regulated. But it holds for commuting
operators and in the zeta-function scheme. Then the
partition function simplifies to
\eqnn{
Z_0={2\sqrt{\pi}\kappa e V\ov m_\gam}
\Big({\det}^\pr(-\lap){\det}^\pr (-\lap+m_ \gamma^2)\Big)
^{-\ha}\,\exp\Big((g_{{\sy 3}}^2+{1\ov 24\pi})S_L\Big).}
We can go further by using the scaling formula for the determinant of $\lap$
\cite{chh4tbv89} and the known result for
the determinant of $\hat\lap$ \cite{Wipf91} which together yield
\eqnl{
{\det}^{\pr\ha}(-\lap)=\tau_0L\vert\eta(\tau)\vert^2
\sqrt{{V\ov \hat V}}\exp\Big(-{1\ov 24\pi}S_L\Big).}{t62}
Thus we obtain the following partition function for the general model
\refs{t1} on curved spaces:
\eqnl{
Z_0=\sqrt{2\pi V}\;{e\ov m_\gam}{1\ov \tau_0\vert\eta(\tau)
\vert^4} \;{1\ov {\det}^{\pr\ha} (-\lap+m_ \gamma^2)}
\,\exp\Big(({1\ov 12\pi}+g_{{\sy 3}}^2)\,S_L\Big).}{t63}
Again we have factored out the partition function
${\cal{N}}_0$ for free auxiliary fields. The formula \refs{t63}  shows
explicitly that in the topologically trivial sector the theory
is equivalent to a theory of free massless
and massive bosons with mass $m_\gam$, even in curved space-time
\cite{Eboli}.\pan
The appearance of $m_\gam$ in \refs{t60} should be interpreted
as {\it renormalization of the electric charge} induced by the
interaction of the auxiliary fields with the fermions. After
summing over all fermion-loops this leads to an effective
coupling between the photons and the $\phi$-field and in turn to
a modified effective mass for the photons in \refs{t60}. In
the limit $ g_{{\sy 2}}\to 0$ this mass tends to the well-known
Schwinger model result, $m_\gam\to e/\sqrt{\pi}$ \cite{chh4ts62}.\pan

We have already mentioned that the
chemical potential coupled to the electric
charge has completely disappeared
from the partition function. This does not come as a
surprise since the only particle in the gauged Thirring model is a
neutral boson. This has no charge
which may couple to the chemical potential. Also, if the partition
function depended on $\mu$ then the
expectation value of the charge would not vanish, in
contradiction to the integrated Gauss law. The result \refs{t63} provides
therefore another test for our result \refs{t36} for the fermionic
determinants with chemical potential. \pan
The final result is also independent
of the chiral and non-chiral twists. The normal twists
have been wiped out by the toron integration.
In fact the chiral twists are equivalent to a chemical potential
and therefore the above remarks concerning the chemical potential apply here
as well.
Did we assume holomorphic factorization for the
fermionic determinant \cite{chh4tdv89} then the partition function
would depend on the chiral twists.\pan
We conclude this subsection by giving the explicit formula
for the partition function on the flat torus.
The zeta-function regularized massive determinant is expressed by
\eqnn{
{\det}^\pr(-\hat\lap+m_ \gamma^2)^\ha ={1\ov m_\gam}
e^{-\ha\zeta^\pr(0)},}
where
\eqnl{
\zeta^\pr(0)= \sum\limits_{n\neq 0}{1\ov \pi L}{
\hat V m_\gam\ov\sqrt{(n,n)}}K_1\big(m_\gam L\sqrt{(n,n)}\big)
-{\hat V m^2_\gam\ov 4\pi},}{t64}
and $(n,n)\es \hat g_{ij}n^in^j$ is the inner product taken
with the reference metric, and the sum is over all $(n^i)\in Z_2$
with the origin excluded. For $g_{\mu\nu}=\delta_{\mu\nu}$, in which
case the partition function has the usual thermodynamical
interpretation, the result reduces to one derived previously
by Ambjorn and Wolfram \cite{ambjorn}. In addition, if $L$ approaches infinity
we
recover a result in \cite{chh4ta87}. The free energy for $\tau_1=0$ and
on flat space simplifies then to
\eqnn{
F=-{1\ov\beta}\log Z = {1\ov 2\beta}\;\zeta^\pr(0).}
with $\zeta^\pr (0)$ from \refs{t64} and the particular choice
for the parameters.
\subsection{Bosonization of the gauged Thirring model}
We pointed out in section $2$ that for $g_{{\sy 1}}\es  g_{{\sy 2}}\es g$
the classical theory \refs{t1} reduces
to the gauged Thirring model. The same is true for the quantized theory on
the torus if in
addition we set $g_{{\sy 0}}\es g$. More precisely, the Hubbard-Stratonovich
transform \cite{chh4th58} of the Thirring model is just the derivative coupling
model \refs{t1} with identical couplings.
In the process of showing that we shall arrive at the Bosonization
formulae for the gauged Thirring model on the curved torus.
We shall see that only the non-harmonic part of
the fermion current can naively be bosonized and that for
this part the rules of the un-gauged model on flat space time \cite{chh4tc75}
need just be covariantized.

For that we calculate the partition function \refs{t54}
in a different order. First we integrate over the auxiliary
fields. To understand the role of $\lam$ and $\phi$ we
introduce sources for them. Thus we study the generating
functional for the correlators of the
auxiliary fields
\eqnn{
Z[\xi,\zeta]=\int \cd (\lam\phi h\psi A_\mu)
e^{\,-S +\int\sqrt{g}\big[\xi\lam+\zeta \phi\big]}.}
Here
\eqnn{
S=-i\int\sqrt{g}\psi^\dagger \di\psi+S_B[g_{{\sy 3}}\es 0]}
is the action of the full theory. $\di$ is the Dirac operator
in \refs{t28} with all couplings set equal and $S_B$ the bosonic
action \refs{t648}.
Since $\lam$ and $\phi$ integrate to zero
(see \refs{52}) we may assume the same property to hold for
the sources. The integration over the auxiliary
fields is Gaussian and yields
\eqnl{
Z={\cal N}_0\!\int\!\cd (\psi A_\mu)\,e^{-S_T}\,
\exp\!\int\sqrt{g} \Big[{-1\ov 4}(\xi{1\ov\lap}\xi+
\zeta{1\ov\lap}\zeta)+{g\ov 2}(\xi{1\ov \lap}j^\mu_{;\mu}+\zeta
{1\ov\lap}j^\mu_{5;\mu})\Big],}{t65}
where
\eqnl{
S_T=-{1\ov 4}\int\sqrt{g}\Big(F_{\mu\nu}F^{\mu\nu}-i\psi^\dagger
\di\psi -{g^2\ov 4}j^\mu j_\mu\Big)}{t66}
is the action of the {\it gauged Thirring model} on curved
space-time and
\eqnl{
{\cal N}_0={V\ov 2\pi {\det}^\pr (-\lap)}}{t67}
comes from the integration over the auxiliary fields.\pan
Let us first consider the partition function, that is set the
sources to zero. Comparing \refs{t65} with \refs{t60} and using
\refs{t62} we easily find
\eqnl{
\int \cd (\psi t)e^{-S_T}=\sqrt{\ha+{g^2\ov 4\pi}}
\;e^{-{1\ov 4}\int F_{\mu\nu}F^{\mu\nu}}
\int\cd \gam\,\delta(\bar\gam)\;e^{-S_\gam},}{t68}
where $\bar\gam$ is the mean field (see \refs{52}) and we used
\refs{t50} and \refs{t58}. The action for the neutral scalar
field $\gam$ is found to be
\eqnn{
S_\gam =\ha\int\sqrt{g} \pa_\mu\gam\pa^\mu\gam - {ie\ov \sqrt{\pi}}
{1\ov\sqrt{1+g^2/2\pi}}\int\sqrt{g}\gam\lap   \varphi.}
Since \refs{t68} holds for any $   \varphi$ (and thus for the
non-harmonic part of any $A_\mu$, because of gauge-invariance)
we read off the following {\it bosonization rules}:
\eqngrl{
j^{\pr\mu}&\longrightarrow &{i\ov\sqrt{\pi}}
{1\ov\sqrt{1+g^2/2\pi}}\eta^{\mu\nu}\pa_\nu\gam}
{j^{\pr\mu}_5&\longrightarrow& -{i\ov\sqrt{\pi}}
{1\ov\sqrt{1+g^2/2\pi}}\pa^\mu\gam,}{t69}
where prime denotes the non-harmonic part of the currents.
Thus, only the non-harmonic parts of the currents can be
bosonized in terms of a single valued scalar field.
To bosonize their harmonic parts one would have to allow for
a scalar field $\gamma$ with winding numbers.
On the infinite plane the harmonic part is not present and
we may leave out the primes in \refs{t69}. If we further assume
space time to be flat we recover the well-known bosonization
rules in \cite{chh4tc75}. What we have shown then, is that for
the gauged model on curved space time the bosonization
rules are just the flat ones properly covariantized and
with the omission of the zero-modes.

Since \refs{t68} holds for any gauge field the current correlators
in the Thirring model are correctly reproduced by the
bosonization rules \refs{t69}. To see that more clearly
we calculate the two-point functions of the auxiliary
fields in the Thirring model (\ref{t65}-\ref{t67}). For
that we differentiate \refs{t65} ($   \varphi$ is treated as
external field) with respect to the sources and find
\eqngrl{
\la \lam(x)\lam(y)\ra& =&\ha G_0(x,y)+
{g^2\ov 4}\int\la G_0(x,u)j^\mu_{;\mu} (u)G_0(y,v)
j^\nu_{;\nu} (v)\rangle_{\sy T}}
{\la \phi(x)\phi(y)\ra& =&\ha G_0(x,y)+
{g^2\ov 4}\int \la G_0(x,u)j^\mu_{5;\mu} (u)G_0(y,v)
j^\nu_{5;\nu}(v)\ra_{\sy T},}{t70}
where $G_0$ is the free massless Green-function
in curved space-time and the integrations are over the variables
$u$ and $v$ with the invariant measure on the curved torus.
Here $\la\dots\ra_{\sy T}$ are vacuum expectation values
in the Thirring model \refs{t66}.
Alternatively we can calculate these expectation values
from \refs{t56} and \refs{t59}, where the fermionic integration has
been performed and find
\eqngr{
\la \lam(x)\lam(y)\ra &=&\ha G_0(x,y)}
{\la \phi(x)\phi(y)\ra &=&{\pi m_ \gamma^2\ov 2 e^2}\,
G_0(x,y)+{m_ \gamma^2\ov 2}\Big(1-{\pi m_ \gamma^2\ov e^2}\Big)
   \varphi (x)   \varphi (y).}
Comparing this with the result \refs{t70} we see at once that
\begin{eqnarray}
\int\la G_0(x,u) j^\mu_{;\mu} (u)G_0(y,v)
j^\nu_{;\nu} (v)\ra_T &=&0\\
\int\la G_0(x,u)j^\mu_{5;\mu} (u)G_0(y,v)
j^\nu_{5;\nu} (v)\ra_T &=&{m_ \gamma^2\ov e^2}\Big(
m_ \gamma^2   \varphi(x)   \varphi(y)-G_0(x,y)\Big).\nonumber
\end{eqnarray}
These correlators express the gauge invariance and the
axial anomaly $\la j^\mu_{5;\mu}\ra\es -m_\gam \lap   \varphi$
in the gauged Thirring model. They can be correctly reproduced
with the bosonization rules \refs{t69}. They are not reproduced
with the ones derived for the un-gauged model \cite{chh4tc75}.

\subsection{Chiral condensate}
The chiral condensate is an order parameter for the chiral symmetry breaking.
However, on the torus its expectation value, whose temperature- and curvature
dependence we will here compute would vanish if topologically
non-trivial gauge field configurations were absent. There is a useful
classification of the gauge configurations
corresponding to the number of fermionic
zero modes they give rise to. If we let $k\es n_+-n_{-}$, where
$n_\pm$ counts the number of
zero-modes with positive/negative chirality, then we have
\eqnl{
k={1\over 2\pi}\int d^2x \,\gamma_5a_1(\di^2,x) ={1\over 4\pi}\int\sqrt{g}
d^2x \,\eta^{\mu\nu}F_{\mu\nu}\equiv {1\over
2\pi}\Phi,}{indexa}
which establishes a relation between the number of fermionic zero
modes (or, more precisely the number of zero modes with positive chirality
minus the number of those with negative chirality) and the first Chern
character of the bundle. Also from \refs{indexa} one
immediately concludes that the flux must be quantized in integer
multiples of $2\pi$. This is really a consequence
of the single valuedness of the fermionic wave function (cocycle
condition).\par
Recalling the decomposition \refs{t520} of the gauge field we now concentrate
on the first term $A^I_\mu$ which is the {\it instanton potential}
giving rise to a non-vanishing quantized flux $\Phi$.
Since $2$-dimensional gauge theories are not scale or Weyl
invariant, as $4$-dimensional ones are, the
instantons on a conformally flat space-time are not identical to the flat
ones \cite{pl2,ket}.
As representative in the $k$-instanton sector we choose the,
up to gauge transformations, {\it unique absolute minimum}
of the Maxwell action in a given topological sector. It has field strength
$e\,E^I=\sqrt{g}\,\Phi/ V$. The corresponding potential can be chosen as
\eqnl{
e A_\mu^I=e\hat A_\mu^I-\Phi\,\eta_\mu^{\;\;\nu}\pa_\nu\chi,
\mtxt{where} e\hat A^I=-{\sqrt{\hat g}\ov\hat V}\Phi\big(x^1,0\big)}{t21}
is the instanton potential on the flat torus with the same flux
but field strength $\sqrt{\hat g}\,\Phi/ \hat V$. The function
$\chi$ is then determined (up to a constant) by
\eqnl{
\sqrt{g}{\Phi\ov V}-\sqrt{\hat g}{\Phi\ov\hat V}=\sqrt{g}\lap\chi.}
{t22}
The solution of this equation is given by
\eqnl{\chi(x)=-{1\ov \hat V}\big({1\ov\lap}e^{-2\si }\big)(x)
={1\ov\hat V}\int d^2y\sqrt{g(y)}G_0(x,y)\,e^{-2\si  (y)},}{t23}
where
\eqnl{
G_0(x,y)=\langle x\vert {1\ov -\lap}\vert y\rangle=\sum_{\lam_n>0}
{\phi_n(x)\phi^\dagger (y)\ov \lam_n}}{t24}
is the Green-function for $-\lap$. In deriving \refs{t23}
we have used that ${1\ov\lap}(\Phi/V)\es 0$ which follows from the
spectral resolution \refs{t24} for the Green function in which the
constant zero mode $\phi_0\es 1/\sqrt{V}$ of $\lap$ is missing.\pan

Our choice for the instanton potential \refs{t21} corresponds
to a particular trivializations of the $U(1)$-bundle over the
torus \cite{chh4tsw92}. In other words, the gauge potentials and fermion
fields at $(x^0,x^1)$ and $(x^0,x^1\!+\!L)$ are necessarily
related by a {\it nontrivial gauge transformation} with winding numbers
\eqngrl{
&&A_{\mu}(x^0,x^1+L)-A_\mu(x^0,x^1)=\pa_\mu\al (x)}
{&&\psi(x^0,x^1+L)=-e^{ie\al (x)}\,e^{2\pi i(\al_1+\beta_1\gf)}\,
\psi(x^0,x^1).}{t26a}
For the choice \refs{t21} we find
\eqnn{
e\al(x)=-{\Phi\ov L}\,x^0.}
Note that $A$ is still periodic in $x^0$ with period $L$ and
$\psi$ still obeys the first boundary condition in \refs{t518}.
To calculate the fermionic {\it zero modes} we use the square of the Dirac
operator
\eqnl{
\di^2=\pmatrix{D_-D_+&0\cr 0&D_+D_-\cr}=
{1\ov \sqrt{g}}D_\mu\sqrt{g}g^{\mu\nu}D_\nu-{1\ov 4}\R
+{e\ov 2}\eta^{\mu\nu}F_{\mu\nu}\gamfive}{dsq}
In a pure instanton and harmonic background ($  \varphi\es \alpha\es 0$) on the
flat torus \refs{dsq} simplifies to
\eqnl{
-\hat\di^2=-\hat g^{\mu\nu}\hat D_\mu\hat D_\nu -{\Phi\ov \hat V}
\gamfive.}{t138}
In other words, $\hat\di^2$ is the same in the left- and right-handed
sectors, up to the constant $2\Phi/\hat V$.
Furthermore this operator commutes with the time translations
which leads to the following ansatz for the zero-modes
\eqnn{
\tilde\chi_p=e^{2\pi ic_px^0/L}\,e^{2\pi iH_1x^1/L}\;\xi_p(x^1)
\,e_+,\quad c_p=\ha +p,}
where we have assumed $k> 0$. The choice of $c_p$ is
dictated by the time-like boundary conditions in \refs{t518}.
Inserting this ansatz into the zero mode equation
$\hat\di^2\tilde\chi_p=0$ yields
\eqngr{
&&\big(|\tau|^2{d^2\ov dy^2}-{\Phi^2\ov L^4}y^2-
2i\tau_1{\Phi\ov L^2}y{d\ov dy}-i\tau{\Phi\ov L^2}\big)
\xi_p =0,}
{&& \mtxt{where} y=x^1+{L\ov k}(c_p-H_0).}
This is just the differential equation for the ground state
of a generalized harmonic oscillator to which it reduces for
$\tau=i\tau_0$. The solution is given by
\eqnn{
\xi_p=\exp\Big[-{\Phi \ov {2i \bar\tau  L^2}}
\big\{x^1+{L\ov k}(c_p-H_0)\big\}^2\Big].}
These functions do not obey the boundary condition \refs{t26a}, but
the correct eigenmodes can be constructed as superpositions of
them. For that we observe that
\eqnn{
\tilde\chi_p(x^0,x^1\!+\!L)=e^{-i\Phi x^0/\beta}\,e^{2i\pi H_1}\;
\tilde\chi_{p+k}(x^0,x^1)}
so that the sums
\eqnl{
\hat\psi^p_0={(2k\tau_0)^{1\ov 4}\ov \sqrt{|\tau|\hat V}}e^{\pi\mu_0^2\ov
k\tau_0}\,e^{2\pi i(H_0-\al_0-\ha)\beta_1}\,
\sum_{n\in Z}\,e^{-2i\pi (n+p/k) (\ha-H_1)}
\tilde\chi_{p+ n k}\,e_+,}{t39}
where $p\es 1,\dots,k$, obey the boundary conditions and thus are
the $k$ required zero-modes. Indeed, since $(i\hat{\di})^2$ in non-negative
there are no zero modes with negative chirality because of \refs{t138}.
With \refs{indexa} we conclude then that there are exactly $k$ zero modes
with positive chirality.
Modes with different $p$ in \refs{t39} are orthogonal to each other
and the overall factor normalized them to one,
so that the system \refs{t39} forms an orthonormal basis of
the zero-mode subspace. For $k\!<\!0$ the zero-modes are the
same if one replaces $e_+$ by $e_-$.\pan
To compute the fermionic determinant in a given topological sector we
again introduce the one-parameter family of Dirac operators
\eqnl{
\di_\tau={\hat g^{1/2}\ov g^{1/2}_\tau}e^{\tau f^\dagger}\,\hat\di\,
e^{\tau f},\qquad\hat\di= \hat\gamma^\mu\Big(\pa_\mu+i\hat\omega_\mu
-ie \hat A^I_\mu-{2\pi i\ov L}[H_\mu+\mu_\mu]\Big),}{t39a}
which interpolates between $\hat\di$ and $\di$,
similarly as in \refs{t539a}. But now
\eqnn{
f=-iF+\gam_5(G+\Phi\chi)+\ha\sigma,}
with $F$ and $G$ from \refs{t5}, contains an instanton contribution.
Also note that $\hat\di$ contains the instanton part $\hat A^I_\mu$. To compute
it's
determinant we observe that the simple form \refs{t138} of $-\hat\di^2$
allows one to reconstruct its spectrum completely
\cite{chh4tbv89,chh4tsw92}:
\eqnn{
\hat\lam _n^2=\cases{0 & degeneracy $=k$ \cr 2n\Phi/\hat V&
degeneracy $=2k.$}}
The corresponding determinant is \cite{chh4tbv89,chh4tsw92}
\eqnl{
{\det}^\pr (i\hat\di)=\Big({\pi\hat V\ov\Phi}\Big)^{\Phi/4\pi}.}{det8}
\par
To relate the determinants of $\hat\di$ to that of $\di$ we again
integrate the anomaly equation, which now reads
\eqnl{\frac{d\log {\det}^\pr \di_\tau}{d\tau}=\int d^2x \sqrt{g_\tau}
\Big(f(x)\!+\!f^\dagger (x)-{d\log g_\tau\ov 2d\tau}\Big)
\{\frac{a_1(x,\di_\tau^2)}{4\pi}-P_0(x,\di_\tau^2)\},}{det15}
where, due to the fermionic zero-modes, the projector onto the
zero-mode-subspace,
\eqnl{P_0(x,\di_\tau^2) =
\sum_{pr}\psi_{p0}^{(\tau)}(x){\cal N}^{-1}_{pr}(\tau)\,(\chi_r^{(\tau)})
^\dagger(x)\mtxt{,}{\cal N}_{pr}(\tau)=\big(\chi_{p0}^{(\tau)},
\psi_{r0}^{(\tau)}\big)}{2.5}
appeared. For the deformed operator ${\di}^2_\tau$ the first
Seeley-deWitt coefficient is
\eqnl{
a_1^\tau=-{1\ov 12}R^\tau+\gamfive \tau\triangle^\tau G
+{1\ov \sqrt{g^\tau}}\,\Big[(1-\tau)\sqrt{\hat g}\,
{\Phi\ov \hat V}
+\tau\sqrt{g}\,{\Phi\ov V}\Big]\gamfive.}{t41}
Integrating w.r.t. $\tau$ \cite{chh4tbv89} one ends up with the following
formula for the determinant in
arbitrary background gravitational and gauge fields:
\eqngrl{
&&{\det}^\pr i\di = \det{{\cal N}_\psi
\ov \hat{\cal N}_\psi}\;{\det}^\pr (i\hat\di)
\exp\Big({S_L\ov 24\pi}+{1\ov 2\pi}\int\sqrt{\hat g}
G\hat\lap G\Big)}
{&&\qquad\quad\cdot\exp\Big({2k\ov V}\int\sqrt{g}G+{\Phi^2\ov 2\pi\hat V}\int
\sqrt{\hat g}\chi\Big).}{t43}
In deriving this result we used
that $\int \sqrt{g}\chi\es 0$.\par
Now we are ready to compute the chiral condensate $\la \psi^\dagger P_+
\psi\ra$.
Observing that the fermionic Green's function anti-commutes with $\gamfive$
one sees at once that only configuration supporting one
fermionic zero-mode with positive chirality contribute to the
chiral condensate
\eqnn{
\la \psi^\dagger P_+\psi\ra=-{J\ov Z_0}
{\delta^2\ov \delta \eta_+(x)\delta\bar\eta_+(x)}\int \cd (\dots)
Z_F[0,0]\,e^{-S_B},}
where $\eta_+\es P_+\eta$. Earlier we have
seen that these are the gauge fields with flux $\Phi\es 2\pi$
or instanton number $k\es 1$. Thus the condensate becomes
\eqnl{
\la \psi^\dagger P_+\psi\ra=-{J\ov Z_0}\sqrt{\hat V\ov 2}
\int\cd (.)\, \psi^\dagger_{0}(x)\psi_{0}(x)
\exp(.)\,e^{-S_B[k=1]},}{t71}
where $\exp(\dots)$ stands for the exponentials in \refs{t43}.
First we integrate
over the toron field $t$. The $t$-dependence enters only
through the zero mode and more specifically $\hat\psi_{0}$ in
\refs{t39} with $p\es 1$. Using the series
representation for the theta functions one finds
\eqnl{
\int d^2 t\,\hat\psi_{0}^\dagger (x)\hat\psi_{0}(x)=
{1\ov \hat V}.}{t72}
Note that the result does not depend on the chemical potential
similarly as in our calculation of the partition function.
To continue we observe that the term $\int \sqrt{g}G$ in
$\exp(\dots)$ vanishes because of our conditions \refs{t49}
and \refs{52} on the fields $\varphi$ and $\phi$.
Furthermore $S_B[k=1]=S_B[k=0]+{2\pi^2\ov e^2V}$.
The remaining functional integrals are performed similarly as
those leading to the partition function and we end up with the
following formula for the condensate
\eqnl{
\la \psi^\dagger P_+\psi\ra=
\sqrt{\tau_0\ov \hat V}\,\vert \eta (\tau)\vert^2
e^{-2\pi^2/e^2V+2\pi/\hat V\int\sqrt{\hat g}\chi}
\Big\la e^{-2(g\phi+e   \varphi)(x)-\si (x)}
\Big\ra_{\phi   \varphi}.}{t73}
The expectation value is evaluated with
\eqnn{
S_{eff}=\int\sqrt{g}\Big[\ha   \varphi (\lap^2-{e^2\ov\pi}\lap)   \varphi
-{e^2\ov\pi m_ \gamma^2}\phi\lap\phi-{e g_{{\sy 2}}\ov\pi}\phi\lap   \varphi
\Big].}
A formal calculation of the resulting Gaussian integrals yields
\eqngrl{
\la \psi^\dagger P_+\psi\ra&=&
\sqrt{\tau_0\ov \hat V}\vert \eta (\tau)\vert^2 e^{-2\pi^2/e^2V
+2\pi/\hat V\int\sqrt{\hat g}\chi}\,
e^{-\si (x)-2\Phi\chi(x)}}
{&\cdot&\exp\big[{2\pi^2m_ \gamma^4\ov e^2}\,K(x,x)\big]
\exp\big[{2\pi  g_{{\sy 2}}^2\ov 2\pi+ g_{{\sy 2}}^2}\,G_0(x,x)\big],}{t74}
where
\eqnl{
K(x,y)=\la x\vert{1\ov \lap^2-m_ \gamma^2\lap}\vert y\ra
={1\ov m_ \gamma^2}\big(G_0(x,y)-G_{m_\gam}(x,y)\big)}{ta74}
and $G_{m},G_0$ are the massive and massless Green-functions.
Here we encounter ultra-violet divergences since
$G_0(x,y)$ is logarithmically divergent when $x$ tends to $y$.
To extract a finite answer we need to renormalize the operator
$\exp(\al \phi)$ as explained in section $3.5$. This wave function
renormalization is equivalent
to the renormalization of the fermion field in the Thirring model
and thus is very much expected \cite{chh4tc75,chh4tsg86}. The flat
Green's function on the torus
\eqnn{\hat G_0(x,y)=-\frac{1}{4\pi}
\big|\frac{1}{\eta(\tau)}\Theta\Big[{\ha+{\xi^0\ov L}\atop
\ha+{\xi^1\ov L}}\Big](0,\tau)\big|^2=-\frac{1}{4\pi}
\big|\frac{1}{\eta(\tau)}e^{i\pi\tau (\xi^0/L)^2}\Theta_1
\big({\tau \xi^0+\xi^1\ov L},\tau\big)\big|^2,}
where $\xi\es x-y$, possesses the logarithmic short
distance singularity
\eqnl{
\hat G_0(x,y)=-{1\ov 4\pi}\log {\hat g_{\mu\nu}\xi^\mu\xi^\nu
\ov \hat V}-{1\ov 4\pi}\log \big(4\pi^2\tau_0
\vert\eta(\tau)\vert^4\big)+O(\xi).}{t75}
Furthermore
\eqnn{
G_0(x,y)\sim \hat G_0(x,y)+2\chi (x)-{1\ov\hat V}
\int\sqrt{\hat g}\chi+O(\xi).}
With the prescription explained in section $3.5$ we find that
on the flat torus $\hat G_0^{reg}$ has now the finite coincidence limit
\eqnl{\hat G_0^{reg}(x,x)=
-{1\ov 4\pi}\log \Big({4\pi^2\tau_0\vert\eta(\tau)\vert^4\ov
\mu^2\hat V}\Big).}{t78}
To determine the chiral condensate we also need to determine
$K(x,y)$ on the diagonal. In a first step we shall obtain
it for the flat torus. Its $\si $-dependence is then determined
in a second step. For $\si \es 0$ and $\tau\es i\tau_0$
the Green function $\hat K$ has been computed in \cite{chh4tsw92}.
The generalization to arbitrary $\tau$ is found to be
\eqngrl{
m_ \gamma^2 \hat K(x,x)&=&-{1\ov 2m_\gam L
\tau_0}\coth\big({\pi\tau_0 a\ov \vert\tau
\vert^2}\big)+{1\ov m_ \gamma^2\hat V}}{
&+&{1\ov 2\pi}
\Big(-\log\vert\eta\big({-1\ov\tau}\big)\vert^2+
F(L,\tau)-H(L,\tau)\Big),}{t79}
where we introduced the dimensionless constant
$a=Lm_\gam\vert\tau\vert/2\pi$ and the functions
\eqngrl{
F(L,\tau)&=&\sum_{n>0}\Big[{1\ov n}-{1\ov \sqrt{n^2+a^2}}\Big]}
{H(L,\tau)&=&\sum_{n>0}{1\ov\sqrt{n^2+a^2}}\Big[{1\ov e^{-2\pi
iz_+(n)}-1}+{1\ov e^{2\pi i z_-(n)}-1}\Big].}{t80}
We used the abbreviations
\eqnl{
z_\pm={1\ov \vert\tau\vert^2}\big(n\tau_1\pm i\tau_0
\sqrt{n^2+a^2}\big).}{t81}
Substituting \refs{t79} and \refs{t78} into \refs{t74} with
$\si \es 0$ we obtain the following {\it exact formula for the
chiral condensate} on the torus with flat metric $\hat g_{\mu\nu}$:
\eqngrl{
\la\psi^\dagger P_+\psi\ra_{\hat g}&=&{1\ov L\vert\tau\vert}\,
\Big({m_\gam L\vert\tau\vert\ov 2\pi}\Big)^{ g_{{\sy 2}}^2\ov 2\pi+
g_{{\sy 2}}^2}
\exp\Big({\pi^2 m_\gam\ov e^2 L\tau_0}\coth{Lm_\gam\tau_0\ov 2
\vert\tau\vert}\Big)}
{&\cdot&\exp\Big[{\pi m_ \gamma^2\ov e^2}\Big(
F(L,\tau)-H(L,\tau)\Big)\Big],}{t82}
where we used that on the flat torus $\chi\es 0$ and $V\es\hat V$.
Furthermore, we identified $\mu$ with the natural mass scale
$m_\gam$ of the theory.

To extract the finite temperature behaviour of the chiral condensate
we take $\tau\es i\beta/L$ where $\beta\es 1/T$
is the inverse temperature. In the
thermodynamic limit $L\to \infty$. Then $\coth(\dots)\to 1$,
$\;H\to 0$ and the expression for the chiral condensate
simplifies to
\eqnl{
\la\psi^\dagger P_+\psi\ra_{\beta}=-
T\Big({m_\gam\ov 2\pi T}\Big)^{ g_{{\sy 2}}^2\ov 2\pi+ g_{{\sy 2}}^2}
\exp\Big[-{\pi^2m_\gam\ov e^2}T+{2\pi \ov 2\pi+ g_{{\sy 2}}^2}F\Big].}{t83}
Using
\eqnn{
F(\beta)\to \gam+\log{a\ov 2}+{1\ov 2 a}\quad\mtxt{for} a\to\infty,}
where $\gam\es 0.57721\dots$ is the Euler number, we obtain the
{\it zero temperature limit}\eqnl{
\la\psi^\dagger P_+\psi\ra=
-{m_\gam\ov 4\pi}\,2^{ g_{{\sy 2}}^2/(2\pi+ g_{{\sy 2}}^2)}\,
\exp\Big({2\pi\ov 2\pi+ g_{{\sy 2}}^2}\gamma\Big)\quad\mtxt{for}
T\to 0.}{t84}
For temperatures large compared to the induced photon mass $F$
vanishes. Thus we obtain the {\it high temperature behaviour}
\eqnl{
\la\psi^\dagger P_+\psi\ra_{T}=
-T\Big({m_\gam\ov 2\pi T}\Big)^{ g_{{\sy 2}}^2\ov 2\pi+ g_{{\sy 2}}^2}\,
\exp\Big(-{\pi^2 m_\gam\ov e^2}T\Big)\mtxt{for} T\to\infty}{t85}
It is instructive to discuss the various limiting cases.
For all $g_i\es 0$, i.e. the Schwinger model limit, the
exact result \refs{t83} simplifies to
\eqnl{
\la\psi^\dagger P_+\psi\ra_T=
-T\,e^{-{\pi\ov m_\gam}T+F(\beta)}\longrightarrow
\cases{-{m_\gam\ov 4\pi}e^\gam &$\quad T\to 0$\cr
-T\,e^{-\pi T/m_\gam}&$\quad T\to\infty$,\cr}}{t86}
where now $m^2_\gam=e^2/\pi$ is the induced photon mass in the
Schwinger model. This formula for the temperature dependence
of the chiral condensate in $QED_2$ agrees with the earlier
results in \cite{chh4tsw92}.\par
Next we wish to investigate how the self interaction of the
fermions affect the breaking. For large coupling $ g_{{\sy 2}}$
and fixed temperature the exponent in \refs{t83} vanishes
so that
\eqnn{
\la\psi^\dagger P_+\psi\ra_T\sim {1\ov \sqrt{2\pi+ g_{{\sy 2}}^2}}\quad
\mtxt{for}T\quad\hbox{fixed},\quad  g_{{\sy 2}}\to\infty.}
Hence, for very large current-current coupling the chiral
condensate vanishes. Or in other words, the electromagnetic
interaction which is responsible for the chiral condensate, is
shielded by the pseudo scalar-fermion interaction.\pan
For intermediate temperature and coupling $ g_{{\sy 2}}$ we must retreat
to numerical evaluations of the sums defining the chiral condensate
in \refs{t83}. The numerical results are depicted in Fig. 1

\iffigs
 \PSGraphic{/home/ivo/diss/chir33.eps}{1.0}
 \vspace{-2.5cm}

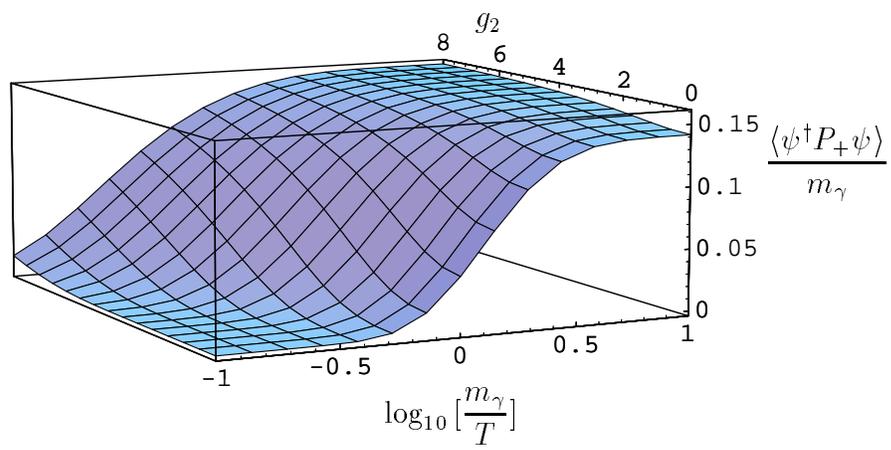
 \begin{center}{Fig. 1: The chiral condensate as a function of the temperature
 and the coupling constant $ g_{{\sy 2}}$.}
 \end{center}
 \vspace{-7.9cm}
 $$\hspace{3cm} g_{{\sy 2}}$$
 \vspace{.5cm}
 $$\hspace{12cm}\frac{\langle\psi^{\dagger}
 P_+\psi\rangle}{m_\gamma}$$
 \vspace{2.1cm}
 $$\hspace{2cm} \log_{10}\big[\frac{m_\gamma}{T}\big]$$
 \vspace{2cm}
 \else
 \message{No figures will be included. See TeX file for more
 information.}
\fi
\par

How does the gravitational field affect
the chiral condensate? To answer this question we need to know the
massive Green's function, entering in (\ref{ta74}), for non-trivial
gravitational fields (for simplicity we assume $T\es 0$). Let us first
consider  a space with constant negative curvature. Then $G_{m_\gamma}$ has
been computed explicitly in \cite{Davies77}. Here we only need its short
distance expansion, given by
\begin{equation}
G_{m_\gamma}(x,y)=-\frac{1}{4\pi}\big\{2\gamma+
\log\big(\frac{-s^2{\cal{R}}}{8}\big)+\psi(\ha+\alpha)
+\psi(\ha-\alpha)+O(s^2)\big\},\label{GmR}
\end{equation}
where $\alpha^2\es\frac{1}{4}+\frac{2m_\gamma^2}{{\cal{R}}}$ and $\psi(z)$
is the Digamma function. Substituting (\ref{GmR}) into (\ref{ta74}) we
end up with the exact formula for the {\it chiral condensate for
constant curvature}
\eqnl{\!\!\!
\langle\psi^{\dagger} P_+\psi\rangle_{\cal{R}}=\langle\psi^{\dagger}
P_+\psi\rangle_{{\cal{R}}=0}\cdot
\exp\Big[\frac{\pi}{2e^2}m_\gamma^2\big\{
\log\big(\frac{-{\cal{R}}}{2m_\gamma^2}\big)
+\psi(\ha\!+\!\alpha)+\psi(\ha\!-\!\alpha)\big\}\Big].}{chirR}
The asymptotic expansions for {\it large-and small
curvature} are easily worked out inserting the corresponding
expansions for the Digamma function \cite{Abr}. We find
\eqnl{
\langle\psi^{\dagger} P_+\psi\rangle_{\cal{R}}=\la\psi^{\dagger}
P_+\psi\rangle_{{\cal{R}}=0}\cdot
\exp\Big[\frac{\pi}{12e^2}{\cal{R}}\Big]\mtxt{for}
\frac{|{\cal{R}}|}{e^2}\ll 1}{smallR}
and
\eqnl{
\la\psi^{\dagger} P_+\psi\rangle_{\cal{R}}=\la\psi^{\dagger}
P_+\psi\rangle_{{\cal{R}}=0}\cdot
\big(\frac{{\cal{R}}}{2m_\gamma^2}\big)^{\frac{\pi}{2\pi+ g_{{\sy 2}}^2}}
\exp\Big[\frac{\pi}{4e^2}
{\cal{R}}- \frac{\pi m_\gamma^2}{4e^2}\gamma\Big]\mtxt{for}
\frac{|{\cal{R}}|}{e^2}\gg 1.}{bigR}
Hence the chiral
condensate decays exponentially for large curvature analogous to the
high temperature behaviour. However, the pseudo-scalars do not suppress
the effect of the curvature in contrast to (\ref{t85}). Comparing the
exponentials in
(\ref{bigR}) to (\ref{t85}) we may define the curvature induced
effective temperature as
\begin{equation}
T_{eff}={-{\cal R}\ov {4\pi m_\gamma}}.
\end{equation}
In passing we note that if we compare the prefactors, rather than the
exponentials, we would write
\begin{equation}
T_{eff}={(-{\cal R})^\ha\ov {4\pi\sqrt{2}}}.
\end{equation}
The latter identification actually coincides (up to factor of $2$) with the
Hawking temperature of free scalars in de Sitter space \cite{chh4tbd82}.
The correct identification involves the (dynamical) mass of
the gauge field and is therefore not universal. From this observation
we learn that the temperature associated with curvature depends
on the matter content. Note finally that the non-minimal coupling
($g_{{\sy 3}}$) has no
effect on the chiral condensate. In Fig. 2 we have
plotted the chiral condensate for arbitrary constant values of the
curvature.\par

\iffigs
 \PSGraphic{/home/ivo/diss/chir22.eps}{1.0}
 \vspace{-2.5cm}
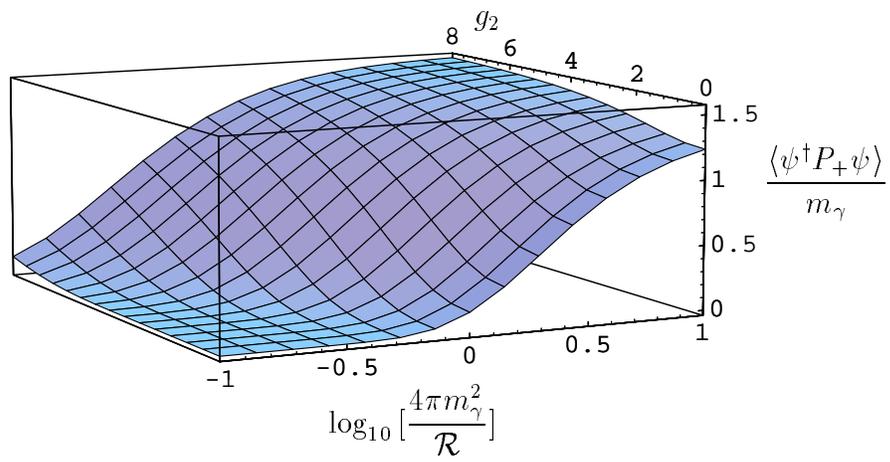
 \begin{center}{Fig. 2: The chiral condensate as a function of the curvature
 and the coupling constant $ g_{{\sy 2}}$.}
 \end{center}
 \vspace{-7.8   cm}
 $$\hspace{3cm} g_{{\sy 2}}$$
 \vspace{.8cm}
 $$\hspace{12cm}\frac{\la\psi^{\dagger}
 P_+\psi\rangle}{m_\gamma}$$
 \vspace{2.2cm}
 $$\hspace{1cm} \log_{10}\big[\frac{4\pi m_\gamma^2}{-{\cal{R}}}\big]$$
 \vspace{1.5cm}
 \else
 \message{No figures will be included. See TeX file for more
 information.}
\fi

\par
For gravitational backgrounds with non-constant curvature we have to
refer to perturbative methods for the calculation of the massive
Green's function. Again we only need the short distance
expansion of $G_{m_\gamma}$. For geodesic distances $s$ small
compared to $m_\gamma^{-1}$ the massive
Green's function may be approximated by the Seeley-DeWitt expansion
\cite{chh4tcr76}
\begin{equation}
G_m(x,y)\sim {1\ov 4i}\sum_{j=0}^\infty a_j(x,y)
\big(-{\pa\ov \pa m^2}\big)^j\,H_0^{(2)}(ms),\label{H0}
\end{equation}
where $H_0^{(2)}$ is
the Hankel function of the second kind and order zero. In particular
$$
H_0^{(2)}(z)\to {2\ov i\pi}\big[\log{z\ov 2}+\gamma\big]
\mtxt{for} z\to 0.
$$
Inserting (\ref{H0}) into (\ref{ta74}) we end up with the following
expansion for the {\it chiral condensate in an arbitrary background}
\begin{equation}
\la\psi^{\dagger} P_+\psi\rangle_{{\cal R}}=\la\psi^\dagger
P_+\psi\rangle_{{{\cal R}}=0}\cdot
\exp\Big[-\frac{\pi}{2} \big({ m_\gamma\ov e}\big)^2 \sum_1^\infty
a_j(x){(j-1)!\ov m^{2j}}\Big],\label{chirS}
\end{equation}
where we have used that $a_0(x)\es1$. The first order
contribution involves $a_1(x)\es -{1\over 6}{\cal R}$ and reproduces
the asymptotic
behaviour (\ref{smallR}). Higher order contributions must be taken
into account to uncover the effect of variable curvature. For this one
has to substitute is the
corresponding Seeley DeWitt coefficients $a_j$ into (\ref{chirS}). These
have been computed up to $j\es 5$ \cite{Avramidi}.\par

\section{Conclusions}
In this paper we have elaborated on various features of the Thirring model as
well as some of its extensions. In particular we found the dependence of the
partition function on the chemical potential and the non-trivial boundary
conditions for the fermions on the torus. For that a careful analysis of
fermionic determinants has been crucial. We have found
that the familiar chiral anomaly of the UV-regularized two point
function is also seen in the IR-sector as a breakdown of holomorphic
factorization. This fact, which has not been properly taken into account
previously, together with the presence of harmonic contributions to the
current, leads to a modification of the equation of state due
to the current-current interaction. We believe that our results could
also be obtained in the bosonized theory, provided the usual bosonization
rules are modified to include scalar fields with winding numbers, i.e.
scalar fields with values in a compactified target space.\pan
Furthermore, we have deformed the conformal structure by allowing
for different couplings in the transversal- and the longitudinal
parts of the current-current interaction. This does not change
the Virasoro- and Kac-Moody algebra, but modifies the conformal
weights of the primaries and in particular of the fermionic fields.
Not all values of the coupling constants belong to physical theories, since
positivity of the scalar product imposes certain restrictions on them.
Our approach allows
also for a non-minimal coupling of the longitudinal sector to gravity. While
such a coupling may seem to be ad-hoc we gave some arguments that it might
arise naturally when quantizing fermions in presence of a a background
charge. We find that the central charge of the Virasoro algebra is
sensitive to the non-minimal coupling. In particular $c<1$ occurs
for certain values of the coupling constant. However, we have not been
able to derive constraints on
this extra coupling without referring to the result by Friedan, Qiu
and Shenker. We believe that an independent derivation of their result
within a fermionic model would be most welcome. We have also
established that the central charge controls the finite size effects only
for a particular treatment of the zero-modes of the auxiliary fields which
is equivalent to an average over charges at infinity.\pan
Finally we have considered the {\it gauged Thirring model} in curved
space-time.
We find that the partition function is independent of vectorial as well as
chiral twists and the chemical potential. This
result, which technically is due to the harmonic contributions to the
gauge-fields, is in fact expected as a consequence of Gauss's law.
Furthermore, using the (probably not so obvious) factorization property
of the zeta-function regularized determinants of commuting operators
we find that the partition function can be expressed completely
in terms of a single massive scalar field. The gauged Thirring model
shows a chiral symmetry breaking which originates in the existence of
fermionic zero-modes and thus in configurations with
winding number (instantons). We have obtained explicit expressions
for these instantons as well as the expectation value of the chiral
condensate as a function of temperature and curvature. The condensate
is exponentially suppressed for high temperatures and/or big curvature
which is interpreted as an almost restoration of the chiral symmetry under
these extreme conditions. Although temperature and curvature have
qualitatively the same effect they cannot be identified. In particular the
identification with the Hawking temperature for free scalar fields in de
Sitter space does not hold in the present situation. It follows from general
arguments that the chiral symmetry can not be restored for any finite
temperature or curvature so an exponential suppression is most we can expect.
In fact, it has been argued earlier, that the axial $U(1)$-symmetry in $4$
dimensional $QCD$ also shows an almost restoration as a function of the
temperature \cite{shur}. Our results on the curvature dependence could
motivate a
corresponding investigation in $QCD$. Finally we note that the chiral
condensate is linearly suppressed for large current-current couplings.

\paragraph{Acknowledgments:} This work has been partially supported by the
Swiss National Science Foundation and the ETH-Z\"urich,
where part of the research leading to the present results has been done.
We wish to thank Arne Dettki for his collaboration at the
beginning of this work and J. Fr\"ohlich, K. Gawedzky,
D. O'Connor and C. Nash for helpful discussions.\par

\begin{appendix}

\section{Conventions and Variational Formulae}
Our conventions for the metric and curvature agree with those of Birrell and
Davies \cite{chh4tbd82}. We use the chiral representation
$\hat\gam^{{\sy 0}}_{{\sy M}}=\sigma_1,
\hat\gam^{{\sy 1}}_{{\sy M}}=i\sigma_2$
for flat space with Lorentzian signature and $\hat\gam^{{\sy 0}}_{{\sy E}}=
\sigma_1,\hat\gam^{{\sy 1}}_{{\sy E}}=-\sigma_2$ in Euclidean space.
Furthermore $\hat\gam_5\es\gamfive=\sigma_3$.\pan
In what follows we derive some variational formulae used in the text.
Here $D_\mu $ denotes the space-time and Lorentz
covariant derivative.\pan
Using the definition of the Christoffel symbols it is straightforward
to show that
\eqngrrl{
&&\quad\delta g_{\mu\nu}=\delta e_\mu^{\;\,a}e _{\nu a}+
e_\mu^{\;\,a}\delta e_{\nu a}\quad ;\quad
\delta\sqrt{g}=\ha \sqrt{g} g^{\mu \nu} \delta g_{\mu \nu}}
{&&\delta\gam^\mu=-\gam^\nu e^\mu_{\;\,a}\delta
e_\nu^{\;\,a}\quad ;
\quad\delta\eta_\mu^{\;\;\nu}=\ha (\eta^{\al\nu}\delta g_{\mu\al}
-\eta_\mu^{\;\;\si}g^{\nu\rho}\delta g_{\si\rho})}
{&&\qquad\quad\delta\Gamma^{\al}_{\mu\nu}=
\ha g^{\al \beta }(D_\nu \delta g_{\beta\mu}
+D_\mu\delta g_{\beta \nu} -D_\beta \delta g_{\mu \nu}).}{B.1}
For some formulae related to the variation of the tetrad let
us refer to
\cite{chh4tlm86}
\eqngrl{
&&\delta e^\mu_{\;\;a}=\ha e_{\nu a}\delta g^{\mu\nu}-t_a^{\;\;b}
e^\mu_{\;\,b}\mtxt {;}
\delta e_\mu^{\;\,a}=\ha e^{\nu a}\delta g_{\mu \nu}-t^a_{\;\;b}
e_\mu^{\;\, b},}
{&&\qquad\mtxt{where}
t^a_{\;\;b}=\ha (e^{\nu a}\delta e_{\nu b}-e^\nu_{\;\, b}\delta
e_\nu^{\;\, a}).}{B.2}
In addition we have
\eqnl{
\delta\omega_{\mu ab}=D_\mu t_{ab} - \al _{\mu ab}\;\quad ;
\al_{\mu ab}=\ha e^\al_{\;\,a}e^\beta_{\;\, b}(D_\al\delta g_{\beta \mu}-
D_\beta\delta g_{\al\mu}).}{B.3}
When performing the variation of curvature dependent expressions
we have used the identities
\eqngrl{
& g^{\mu\nu}\delta \R_{\mu\nu}=\omega^{\al}_{\;;\al}\;,\mtxt{where}
\omega^{\al}=g^{\mu \nu} \delta\Gamma^{\al}_{\mu\nu}-g^{\al\nu}
\delta\Gamma^\mu_{\mu\nu}}
{&\mtxt{and}\int\sqrt{g}\,\omega^{\al}A_{\al}=\int\sqrt {g}
\{g^{\al\beta}\nabla_{\mu} A^{\mu}-\nabla^{\al}A^{\beta}\}
\delta g_{\al\beta}\;.}{B.4}
Depending on the topology of space-time, the reference
curvature $\hat \R $ may be
different from zero. In this case it is not possible to express
the conformal angle $\si $ in terms of the curvature scalar.
Nevertheless, to perform variations of $\si$-dependent expressions, the
identity
\eqnl{
\delta (\sqrt{g}\R )=-2\delta(\sqrt{g}\lap\si)}{B.5}
proves to be useful.\pan
Taking the variations of the equations
\eqnl{
\sqrt{g}\Box G(x,y) =-\de(x-y) \mtxt{and}\sqrt{g}\,i\di S(x,y)=
\de (x-y)}{B.6}
for the scalar and fermionic Greens functions we may derive
(up to contact terms) the following variational formulae
\eqngr{
\de G=\int\;\big(-\ha g^{\mu\nu}g^{\al\beta}
+g^{\al\mu}g^{\beta\nu}\big)\pa_{\al} G(x,u)\,\pa_{\beta}
G(u,y)\sqrt{g}\delta g_{\mu\nu}}
{\de S={i\ov 4}\int\;\Big(2S(x,u)\gam^\mu D^\nu
S(u,y)\!-\!D_\al [S(x,u)\gam^\delta\eta_\delta^{\;\,\mu}
\eta^{\nu\al} S(u,y)]\Big)\sqrt{g}\delta g_{\mu\nu}.}
Here all arguments and derivatives which are not made explicit
in the integral refer to the coordinate $u$ over which is
integrated. Finally, we need the following formula for the
variation of the inverse Laplacian
\eqnl{
\delta\left(\ilap f\right) =\ilap\left(\delta f-\delta (\lap )
\ilap f\right)-{1\ov 2V}\int\sqrt{g}g^{\mu\nu}\delta g_{\mu\nu}\ilap f
, }{B.10}
where $V$ is the volume of space-time and $f$ an arbitrary function. To
prove this identity we note that for $f\in (\hbox{Kern}\triangle)^{\bot}$ we
have
\eqnn{\lap\ilap f=f.}
Varying this equation yields
\eqnn{
\lap (\delta \ilap f ) =\delta f-(\delta \lap ) \ilap f}
which may be inverted to give
\eqnl{
\delta \left(\ilap f \right )  = \ilap \left (\delta f
- \delta (\lap )\ilap f \right) + {1\ov V} \int \sqrt{g}
\delta \left(\ilap f \right ) \;.}{B.11}
Varying the identity
\eqnn{{1\ov V} \int \sqrt{g}\ilap f =\; 0}
allows to replace the last term of \refs{B.11} to obtain the required
result \refs{B.10}.

\section{Canonical Approach to the Partition Function}
In this appendix we compute the partition function for massive
Dirac fermions in the canonical formalism. In the limit $m\to 0$
we confirm the result \refs{t36} for the fermionic determinant with
chemical potential in chapter 3.
For massive fermions one cannot consistently impose chirally twisted
boundary conditions. However, from the explicit eigenvalues \refs{t31b}
one sees at once that the chiral twist $\beta_1$ and the
chemical potential are equivalent. One can easily verify that
this equivalence holds also for massless fermions in the canonical approach
and that $\beta_1\sim \mu L/2\pi$. Let us therefore compute the partition
function
\eqnl{
Z(\beta)=Tr\big[e^{-\beta:(H-\mu Q):}\big]}{C.1}
for massive Dirac fermions with chemical potential $\mu$ on a
cylinder with (non chiral) twisted boundary conditions
\eqnl{
\psi(x+L,t)=-e^{-2i\pi\al_1}\psi(x,t).}{C2}
For massive particles it is more convenient to use the Dirac representation
\eqnl{
\gam^0=\sigma_3\quad\gam^1=-i\sigma_2,\quad\gam^5=\gam^0\gam^1
=-\sigma_1.}{C.3}
The Dirac field is expanded in terms of the eigenmodes of the
first quantized Hamiltonian
\eqnl{
h=\pmatrix{m&i\pa_x\cr i\pa_x&-m\cr}}{C.4}
as
\eqnl{
\Psi(x,t)=\sum_n \psi_{n,+}b_n+\sum_n \psi_{n,-}d^\dagger_n,}{C.5a}
where the $\psi_{n,+}$ and $\psi_{n,-}$ are the positive and
negative energy modes,
\begin{eqnarray}
&&\psi_{n,+}=e^{-i\om_nt-i\lam_n x}c_n,\;\;\;
\psi_{n,-}=e^{i\om_nt-i\lam_n x}\gam_1 c_n,\nonumber\\
&& c_n=\big(2\om_n (\om_n+m)L\big)^{-\ha}\pmatrix{\om_n+m\cr
\lam_n\cr}.\label{C.5b}
\end{eqnarray}
The momenta $\lam_n$ and frequencies $\om_n$ are determined by the
boundary condition \refs{C2} to be
\eqnl{
\lam_n={2\pi\ov L}\big(n-\ha-\al_1\big)\mtxt{and} \om_n=\sqrt{m^2+\lam_n^2}.}
{C.5c}
After normal ordering the 'positron' operators with respect to
the Fock vacuum defined by $H$ we find
\eqnl{
(H-\mu Q)=
\sum_n(\om_n-\mu)b^\dagger_nb_n+\sum_n (\om_n+\mu)d^\dagger_n d_n
-\sum_n(\om_n+\mu),}{C.6}
where the last $c$-number term represents the infinite vacuum
contribution which must be regularized.
To do that we employ the zeta function regularization.
That is we define the zeta-function for $s\!>\!1$ by the sum
\eqnn{
\zeta(s)=\sum_n(\om_n+\mu)^{-s},}
which in turn defines an analytic function on the whole complex $s$-plane
up to a simple pole at $s\es 1$. The analytic continuation is constructed
by a Poisson resummation
\eqnl{
\sum_{n}(\om_n+\mu)^{-s}={L^s\ov 2\pi}\sum_{n}F(n),}{C7a}
where
\eqnl{
F(\xi)= e^{2\pi i\xi(\ha-\al_1)}\int
dy\; e^{i\xi y}\big[\sqrt{\tilde m^2+y^2}\ +\tilde\mu\big]^{-s}}{C7b}
and $\tilde m \es Lm$, $\tilde\mu\es L\mu$. Taking the Mellin transform
of \refs{C7b} we find
\begin{eqnarray}
F(\xi)&=& e^{2\pi i\xi(\ha-\al_1)}{1\ov\Gamma(s)}\int dy\ e^{i\xi y}\int
dt\ t^{s-1}e^{-t\sqrt{\tilde m^2+y^2}-t\tilde\mu}\nonumber\\
&=&-{2\ov\Gamma(s)}e^{2\pi i\xi(\ha-\al_1)}\int dt\ t^{s-1}e^{-t\tilde\mu}\
{d\ov dt}K_0(\tilde\mu\sqrt{\xi^2+t^2})\\
&=&{2\tilde m\ov\Gamma(s)}e^{2\pi i\xi(\ha-\al_1)}\int dt\ t^s e^{-t\tilde\mu}
{K_1(\tilde\mu\sqrt{\xi^2+t^2})\ov
\sqrt{\xi^2+t^2}}.\nonumber\label{C.7c}
\end{eqnarray}
$F$ diverges at $\xi\es 0$ since the Kelvin function
$K_1(z)\sim 1/z$ for small $z$. It follows that the $n\es0$
term in \refs{C7a} diverges. This divergence is regularized by subtracting
the ground state energy of the infinite volume system. Indeed,
because of the exponential decay of the Bessel function for large
arguments, only the $n\es 0$ term contributes for infinite volume. So we
find for the regularized sum
\eqnl{
\sum_n(\om_n+\mu)^{-s}={\tilde m L^s\ov\Gamma(s)\pi}\sum_{n\neq 0}\!\int
dt\ e^{2\pi in(\ha-\al_1)} t^s e^{-t\tilde\mu}{K_1(\tilde m\sqrt{n^2+t^2})
\ov \sqrt{n^2+t^2}}.}{C.8}
Now we perform the limit $m\rightarrow 0$. Only the most
singular term in the expansion of the Bessel function contributes,
hence
\begin{eqnarray}
\sum_n(\om_n+\mu)^{-s}&=&{L^s\ov\Gamma(s)\pi}\sum_{n\neq 0} \int
dt\ e^{2\pi in(\ha-\al_1)} t^s e^{-t\tilde\mu}{1\ov (n^2+t^2)}\nonumber\\
&=&{sL^s\ov \pi}\sum_{n\neq 0} e^{2\pi in(\ha-\al_1)}\sqrt{\tilde\mu}
n^{s-\ha}S_{-s-\ha;\ha}(\tilde\mu n),\label{C.9}
\end{eqnarray}
where $S_{a;b}(z)$ is the Lommel function \cite{Grad}. In particular for
$s\es-1$ this function is  $S=1/z$ so that finally
\eqnl{
\sum_n(\om_n+\mu)^{reg}=-{1\ov \pi L}\sum_{n\neq 0}{(-)^n\ov n^2} e^{-2\pi
in\al_1} ={\pi\ov 6L}-{2\pi\ov L}(\al_1-[\al_1\!+\!\ha])^2.}{C.10}
Inserting this into \refs{C.6} then yields the regularized expression
\eqnl{
:H-\mu Q:=\sum_n(\om_n\!-\!\mu)b^\dagger_nb_n +\sum_n (\om_n\!+\!\mu)
b^\dagger_nd_n-{\pi\ov 6L}+{2\pi\ov L}\Big(\al_1-[\al_1\!+\!\ha]\Big)^2.}{C.11}
For small $\mu$ the normal ordering is $\mu$-independent so that
\eqnl{
\la 0\vert :H-\mu Q:\vert 0\rangle=
-{\pi\ov 6L}+{2\pi\ov L}\Big(\al_1-[\al_1+\ha]\Big)^2=
\la 0\vert :H:\vert 0\rangle}{C12}
is independent of $\mu$ and coincides with the Casimir energy
\cite{chh4tkw93}. \par
Let us now compute the partition function. Using \refs{C12} we
easily find
\begin{eqnarray}
Z(\beta)&=&\tr\big[e^{-\beta:(H-\mu Q):}\big]=q^{[\al_1^2-{1\ov
12}]}\nonumber\\
&=&\prod_{n>[\ha+\al_1]}^\infty
(1+q^{(n-\ha-\al_1)}e^{\beta\mu})\prod_{n>-[\ha+\al_1]}^\infty
(1+q^{(n-\ha+\al_1)}e^{\beta\mu})\cdot\nonumber\\
&&\;\prod_{n>[\ha-\al_1]}^\infty
(1+q^{(n-\ha+\al_1)}e^{-\beta\mu})\prod_{n>-[\ha-\al_1]}^\infty
(1+q^{(n-\ha-\al_1)}e^{-\beta\mu})\nonumber\\
&=& {1\over |\eta(\tau)|^2}
\Theta\Big[{-\al_1\atop
i\mu {\beta\ov 2\pi}}\Big](0,\tau)\
\bar\Theta\Big[{ -\al_1   \atop
-i\mu {\beta\ov 2\pi} }\Big](0,\tau),\label{C.13}
\end{eqnarray}
where we have used the product representation of the theta functions
in the last identity and that $q=e^{2\pi i\tau}=e^{-2\pi\beta/L}$.
A non-vanishing chiral twist $\beta_1$ can now be included
by shifting the chemical potential. Thus we have confirmed
the formula \refs{t36} in the text.\par
Note that for $\mu\neq 0$ the zero-temperature limit of the
grand potential is not equal to the vacuum expectation
value of $:\!H-\mu Q\!:$\,. For $\mu\neq 0$ all states up to
the $\mu$-dependent Fermi energy are filled. For example,
for $\om_1<\mu<\om_2$ in the limit $\beta\to\infty$, $\Omega$
reduces to the expectation value of $:H-\mu Q:$ in the one-electron
state.

\end{appendix}


\addcontentsline{toc}{chapter}{Bibliography}
\begin{thebibliography}{99}
\bibitem{chh4tdr84} see e.g. W. Dittrich  and M. Reuter, "Effective
Lagrangians in QED", Lecture notes in Physics, Springer,
Heidelberg, 1984 and references therein.
\bibitem{chh4tt58} W.E. Thirring, {\it Ann. Phys. (NY)} {\bf 3}
(1958), 91; V. Glaser, Nuovo Cim. {\bf 9},
(1958), 1007.
\bibitem{chh4ty87} H. Yokota, Prog. Theor. Phys. {\bf 77} (1987), 1450.
\bibitem{chh4tfp88} D.Z. Freedman and K. Pilch, Phys. Lett. {\bf 213B} (1988),
331; {\it Ann. Phys. (NY)} {\bf 192} (1989), 331;
S. Wu, Comm. Math. Phys. {\bf 124} (1989), 133.
\bibitem{chh4tdv89} C. Destri and J.J. deVega, Phys. Lett. {\bf 223B} (1989),
365.
\bibitem{chh4tbd82} N.D. Birrell and P.C.W. Davies, "Quantum
fields in curved space", Cambridge Univ. Press, 1982.
\bibitem{chh4tgp78} G.W. Gibbons and M.J. Perry, Proc. R. Soc. London {\bf A
358} (1978), 467.
\bibitem{chh4tsw92} I. Sachs and A. Wipf, Helv. Phys. Acta
{\bf 65} (1992), 653.
\bibitem{chh4tls92} H. Leutwyler and V. Smilga, Phys. Rev.
{\bf D46} (1992), 5607.
\bibitem{chh4tj64} K. Johnson, Nuovo Cim. {\bf 20} (1964), 773.
\bibitem{chh4tj88} C. Jayewardena, Helv. Phys. Acta
{\bf 61} (1988), 636; H. Joos, Nucl. Phys. {\bf B17} (Proc. Suppl.) (1990),
704; Helv. Phys. Acta {\bf 63} (1990), 670; K.D. Rothe and  J.A. Swieca, {\it
Ann. Phys (N.Y.)} {\bf 117} (1979), 382; J. Verbaarschot et.al, Nucl. Phys.
{\bf 425} (1994), 553.
\bibitem{chh4tbd78} N.D. Birrell and P.C.W. Davies, Phys. Rev. {\bf D18}
(1978), 4408.
\bibitem{Eboli} O.J.P. Eboli, Phys. Rev. {\bf D36} (1987), 2408.
\bibitem{chh4tg94} K. Gawedzky, "Conformal field theory",
to appear in Birkh{\"a}user, Basel.
\bibitem{denjoe} C. Nash and D. O'Connor, {\it Preprint} DIAS-STP-95-24.
\bibitem{chh4tk68} B. Klaiber, ``Lecture notes in Phys. XA'',
Gordon and Breach, New York, 1968.
\bibitem{chh4tc76} R. Casalbuoni, Nuovo Cim. {\bf A33} (1976),
115
\bibitem{chh4taw83} L. Alvarez-Gaume and E. Witten, Nucl. Phys. {\bf B234}
(1983), 269; H. Leutwyler, Phys. Lett. {\bf 153B} (1985), 65; L. Alvarez-Gaume
et. al, Comm. Math. Phys. {\bf 112} (1987), 503; E. D'Hoker and D.H. Phong,
Nucl. Phys. {\bf B278} (1
986), 226; K. Harada and
I. Tsutsui, Phys. Lett. {\bf B183} (1987), 311.
\bibitem{chh4ta87} A. Actor, Fortschritte der Phys.
{\bf 35} (1987), 793; K. Tsokos, Phys. Lett. {bf 157B} (1985), 413.
\bibitem{chh4tbv89} S. Blau, M. Visser and A. Wipf, Int. J. Mod. Phys. {\bf A4}
(1989), 1467.
\bibitem{Wipf91} S. Blau, M. Visser and A. Wipf, Int. J. Mod. Phys. {\bf A6}
(1991), 5409.
\bibitem{chh4tp87} see e.g. A.M. Polyakov, "Gauge fields
and Strings'', Harwood Academic Publishers, 1987.
\bibitem{WW} C. Wiesendanger and A. Wipf, {\it Ann. of Phys. (NY)}
{\bf 233} (1994),
{\bf 233}, 125.
\bibitem{chh4tkw93} C. Kiefer and A. Wipf, {\it Ann. of Phys. (NY)}
{\bf 236} (1994), 241; S. Iso and H. Murayama, Progr. Theor. Phys.
{\bf 84} (1990), 142.
\bibitem{chh4tlt89} D. L{\"u}st and S. Theisen, "Lectures
on string theory", Lecture notes in physics, Springer, 1989.
\bibitem{chh4tfs89} P. Furlan, G.M. Sotkov and I.T. Todorov, Riv. Nuovo Cim.
{\bf 12} (1989), 1.
\bibitem{chh4trw90} L. O'Raifeartaigh and A. Wipf, Phys. Lett. {\bf 251B}
(1990), 361.
\bibitem{OS} see eg. J. Glimm and A. Jaffe, ``Quantum Physics'', second
edition,
Springer, New York, 1987.
\bibitem{thesis} I. Sachs, PhD. Thesis, ETH-No. 10728 (1994),
{\it unpublished}.
\bibitem{chh4tc88} J.L. Cardy, {\it in} "Fields, Strings and
Statistical Mechanics'', Les Houches, 1988.
\bibitem{chh4tdw92} A. Dettki and A. Wipf, Nucl. Phys.
{\bf B377} (1992), 252.
\bibitem{chh4tc75} S. Coleman, Phys. Rev. {\bf D11} (1975), 2088.
\bibitem{chh4tsg86} A.J. da Silva, M. Gomes and R. K{\"o}berle, Phys. Rev. {\bf
D34}
(1986), 504; M. Gomes and A.J, da Silva, Phys. Rev. {\bf D34}
(1986), 3916.
\bibitem{Banks} see eg. T. Banks, {\it in} Proceedings of the Santa Fe TASI-87,
Vol. II, 1987.
\bibitem{chh4tln93} F. Lenz, H.W.L. Naus and M. Thies, {\it Ann. of Phys. (NY)}
{\bf 233} (1994), 17.
\bibitem{chh4ts93} A.V. Smilga, {\it Ann. of Phys. (NY)} {\bf 234} (1994), 1.
\bibitem{chh4ts79} L. Susskind, Phys. Rev. {\bf D20} (1979), 2610;
N. Weiss, Phys. Rev. {\bf D24} (1981), 475.
\bibitem{chh4ts62} J. Schwinger, Phys. Rev. {\bf 128} (1962), 2425.
\bibitem{ambjorn} J.Ambjorn and S. Wolfram, {\it Ann. of Phys. (NY)} {\bf 147}
(1983), 1.
\bibitem{chh4th58} J. Hubbard, Phys. Rev. Lett. {\bf 3} (1958), 77;
R.L. Stratonovich, Sov. Phys. Dokl. {\bf 2} (1958), 416.
\bibitem{chh4gs93} C. Gattringer and E. Seiler, {\it Ann. of Phys. (NY)} {\bf
233} (1994), 97.
\bibitem{pl2} I. Sachs and A. Wipf, Phys. Lett. {\bf B326} (1994), 105.
\bibitem{ket} S.V. Ketov and O. Lechtenfeld, Phys. Lett. {\bf B353} (1995),
463.
\bibitem{Davies77} T.S. Bunch and P.C.W Davies, Proc. R. Soc.
Lond. {\bf A360} (1978), 117.
\bibitem{Abr} M. Abramowitz and I.S. Stegun, "Handbook of
Mathematical Functions", Dover Publications Inc. NY, 1972.
\bibitem{chh4tcr76} S.M. Christensen, Phys. Rev. {\bf D14} (1976), 2490.
\bibitem{Avramidi} I.G. Avramidi, Phys. Lett. {\bf B238} (1990), 92.
\bibitem{chh4tlm86} H. Leutwyler and S. Mallik, Z. Phys.
{\bf C33} (1986), 205.
\bibitem{shur} E.V. Shuryak, Phys. Lett. {\bf 79B} (1978), 135;
R.D. Pisarski and L.G. Yaffe, Phys. Lett.,
{\bf 97B} (1980), 110.
\bibitem{Grad} I.S. Gradshteyn and I.M. Ryzhik, "Table of Integrals,
Series and Products", Academic, London, 1980.








\end{thebibliography}
\end{document}

\end